\documentclass[ALICE,manyauthors]{cernphprep}
\usepackage[comma,square,numbers,sort&compress]{natbib}
\usepackage{hyperref}
\usepackage{lineno}
\usepackage{enumitem}
\usepackage{xspace}
\usepackage{booktabs}
\usepackage{multirow}
\usepackage{tablefootnote}
\usepackage[T1]{fontenc}
\usepackage{orcidlink}

\begin{document}
%

\newcommand{\pp}           {pp\xspace}
\newcommand{\ppbar}        {\mbox{$\mathrm {p\overline{p}}$}\xspace}
\newcommand{\ccbar}        {\mbox{$\mathrm {c\overline{c}}$}\xspace}
\newcommand{\bbbar}        {\mbox{$\mathrm {b\overline{b}}$}\xspace}
\newcommand{\XeXe}         {\mbox{Xe--Xe}\xspace}
\newcommand{\PbPb}         {\mbox{Pb--Pb}\xspace}
\newcommand{\pA}           {\mbox{pA}\xspace}
\newcommand{\pPb}          {\mbox{p--Pb}\xspace}
\newcommand{\AuAu}         {\mbox{Au--Au}\xspace}
\newcommand{\dAu}          {\mbox{d--Au}\xspace}

\newcommand{\s}            {\ensuremath{\sqrt{s}}\xspace}
\newcommand{\snn}          {\ensuremath{\sqrt{s_{\mathrm{NN}}}}\xspace}
\newcommand{\pte}          {\ensuremath{p_{\rm T,e}}\xspace}
\newcommand{\pt}           {\ensuremath{p_{\rm T}}\xspace}
\newcommand{\meanpt}       {$\langle p_{\mathrm{T}}\rangle$\xspace}
\newcommand{\ycms}         {\ensuremath{y_{\rm CMS}}\xspace}
\newcommand{\ylab}         {\ensuremath{y_{\rm lab}}\xspace}
\newcommand{\etarange}[1]  {\mbox{$\left | \eta \right |\kern-.5em~<~\kern-.2em#1$}}
\newcommand{\etarangee}[1] {\mbox{$\left | \eta_{\rm e} \right |~<~#1$}}
\newcommand{\yrange}[1]    {\mbox{$\left | y \right |~<~#1$}}
\newcommand{\dndy}         {\ensuremath{\mathrm{d}N_\mathrm{ch}/\mathrm{d}y}\xspace}
\newcommand{\dndeta}       {\ensuremath{\mathrm{d}N_\mathrm{ch}/\mathrm{d}\eta}\xspace}
\newcommand{\avdndeta}     {\ensuremath{\langle\dndeta\rangle}\xspace}
\newcommand{\dNdy}         {\ensuremath{\mathrm{d}N_\mathrm{ch}/\mathrm{d}y}\xspace}
\newcommand{\Npart}        {\ensuremath{N_\mathrm{part}}\xspace}
\newcommand{\Ncoll}        {\ensuremath{N_\mathrm{coll}}\xspace}
\newcommand{\TpPb}         {\ensuremath{\langle T_{\rm pPb} \rangle}\xspace}
\newcommand{\dEdx}         {\ensuremath{\textrm{d}E/\textrm{d}x}\xspace}
\newcommand{\RpPb}         {\ensuremath{R_{\rm pPb}}\xspace}
\newcommand{\phiV}         {\ensuremath{\varphi_{\kern.1em\rm V}}}

\newcommand{\nineH}        {$\sqrt{s}~=~0.9$~Te\kern-.1emV\xspace}
\newcommand{\seven}        {$\sqrt{s}~=~7$~Te\kern-.1emV\xspace}
\newcommand{\twoH}         {$\sqrt{s}~=~0.2$~Te\kern-.1emV\xspace}
\newcommand{\twosevensix}  {$\sqrt{s}~=~2.76$~Te\kern-.1emV\xspace}
\newcommand{\five}         {$\sqrt{s}~=~5.02$~Te\kern-.1emV\xspace}
\newcommand{\thirteen}     {$\sqrt{s}~=~13$~Te\kern-.1emV\xspace}
\newcommand{\twosevensixnn}{$\sqrt{s_{\mathrm{NN}}}~=~2.76$~Te\kern-.1emV\xspace}
\newcommand{\srichnn}      {$\sqrt{s_{\mathrm{NN}}}~=~200$~Ge\kern-.1emV\xspace}
\newcommand{\fivenn}       {$\sqrt{s_{\mathrm{NN}}}~=~5.02$~Te\kern-.1emV\xspace}
\newcommand{\LT}           {L{\'e}vy-Tsallis\xspace}
\newcommand{\GeVc}         {Ge\kern-.1emV/$c$\xspace}
\newcommand{\MeVc}         {Me\kern-.1emV/$c$\xspace}
\newcommand{\TeV}          {Te\kern-.1emV\xspace}
\newcommand{\GeV}          {Ge\kern-.1emV\xspace}

\newcommand{\MeV}          {Me\kern-.1emV\xspace}
\newcommand{\GeVcc}        {Ge\kern-.1emV/$c^2$\xspace}
\newcommand{\MeVcc}        {Me\kern-.1emV/$c^2$\xspace}
\newcommand{\lumi}         {\ensuremath{\mathcal{L}}\xspace}
\newcommand{\mee}          {\ensuremath{m_{\rm ee}}\xspace}
\newcommand{\ptee}         {\ensuremath{p_{\rm T, ee}}\xspace}
\newcommand{\meeRange}[2]  {\ensuremath{{#1}< m_{\rm ee} <{#2}}~\GeVcc}
\newcommand{\mllRange}[2]  {\ensuremath{{#1}< m_{\rm ll} <{#2}}~\GeVcc}
\newcommand{\meeMin}[1]    {\ensuremath{m_{\rm ee} >{#1}}~\GeVcc}
\newcommand{\meeMax}[1]    {\ensuremath{m_{\rm ee} <{#1}}~\GeVcc}
\newcommand{\pteeRange}[2] {\ensuremath{{#1}< p_{\rm T,ee} <{#2}}~\GeVc}
\newcommand{\pteeMin}[1]   {\ensuremath{p_{\rm T,ee} >{#1}}~\GeVc}
\newcommand{\pteeMax}[1]   {\ensuremath{p_{\rm T,ee} <{#1}}~\GeVc}
\newcommand{\ptMin}[1]   {\ensuremath{p_{\rm T} >{#1}}~\GeVc}
\newcommand{\ptMax}[1]   {\ensuremath{p_{\rm T} <{#1}}~\GeVc}
\newcommand{\pteMin}[1]   {\ensuremath{p_{\rm T,e} >{#1}}~\GeVc}
\newcommand{\pteMax}[1]   {\ensuremath{p_{\rm T,e} <{#1}}~\GeVc}
\newcommand{\pteRange}[2] {\ensuremath{{#1}< p_{\rm T,e} <{#2}}~\GeVc}
\newcommand{\nSig}[2]      {\ensuremath{\rm{n}\sigma_{#1}^{#2}}}
\newcommand{\ITS}          {\rm{ITS}\xspace}
\newcommand{\TOF}          {\rm{TOF}\xspace}
\newcommand{\ZDC}          {\rm{ZDC}\xspace}
\newcommand{\ZDCs}         {\rm{ZDCs}\xspace}
\newcommand{\ZNA}          {\rm{ZNA}\xspace}
\newcommand{\ZNC}          {\rm{ZNC}\xspace}
\newcommand{\SPD}          {\rm{SPD}\xspace}
\newcommand{\SDD}          {\rm{SDD}\xspace}
\newcommand{\SSD}          {\rm{SSD}\xspace}
\newcommand{\TPC}          {\rm{TPC}\xspace}
\newcommand{\TRD}          {\rm{TRD}\xspace}
\newcommand{\VZERO}        {\rm{V0}\xspace}
\newcommand{\VZEROA}       {\rm{V0A}\xspace}
\newcommand{\VZEROC}       {\rm{V0C}\xspace}
\newcommand{\Vdecay} 	   {\ensuremath{V^{0}}\xspace}

\newcommand{\ee}           {\ensuremath{{\rm e}^{+}{\rm e}^{-}}} 
\newcommand{\llepton}        {\ensuremath{{\rm l}^{+}{\rm l}^{-}}} 
\newcommand{\pip}          {\ensuremath{\pi^{+}}\xspace}
\newcommand{\pim}          {\ensuremath{\pi^{-}}\xspace}
\newcommand{\kap}          {\ensuremath{\rm{K}^{+}}\xspace}
\newcommand{\kam}          {\ensuremath{\rm{K}^{-}}\xspace}
\newcommand{\pbar}         {\ensuremath{\rm\overline{p}}\xspace}
\newcommand{\jpsi}         {\ensuremath{{\rm J}/\psi}\xspace}
\newcommand{\kzero}        {\ensuremath{{\rm K}^{0}_{\rm{S}}}\xspace}
\newcommand{\lmb}          {\ensuremath{\Lambda}\xspace}
\newcommand{\almb}         {\ensuremath{\overline{\Lambda}}\xspace}
\newcommand{\Om}           {\ensuremath{\Omega^-}\xspace}
\newcommand{\Mo}           {\ensuremath{\overline{\Omega}^+}\xspace}
\newcommand{\X}            {\ensuremath{\Xi^-}\xspace}
\newcommand{\Ix}           {\ensuremath{\overline{\Xi}^+}\xspace}
\newcommand{\Xis}          {\ensuremath{\Xi^{\pm}}\xspace}
\newcommand{\Oms}          {\ensuremath{\Omega^{\pm}}\xspace}
\newcommand{\degree}       {\ensuremath{^{\rm o}}\xspace}


\newcommand{\sNN}{\ensuremath{\sqrt{s_{\mathrm{NN}}}}\xspace}
\newcommand{\DCAee}{\ensuremath{\mathrm{DCA_{ee}}}\xspace}

\newcommand{\Jpsi}{\ensuremath{\mathrm{J}/\psi }\xspace}

\newcommand{\qqbar}{\ensuremath{\mathrm {q\bar{q}}}\xspace}
\newcommand{\QQbar}{\ensuremath{\mathrm {Q\overline{Q}}}\xspace}
\newcommand{\DDbar}{\ensuremath{\mathrm {D\overline{D}}}\xspace}
\newcommand{\BBbar}{\ensuremath{\mathrm {B\overline{B}}}\xspace}

\newcommand{\cHc}{\ensuremath{\mathrm{c \rightarrow H_{c}}}\xspace}
\newcommand{\Hce}{\ensuremath{\mathrm{H_{c} \rightarrow e}}\xspace}

\begin{titlepage}
\PHyear{2020}       
\PHnumber{081}      
\PHdate{19 May}  

\title{Dielectron production in proton--proton and proton--lead collisions at $\mathbf{\sqrt{{\textit s}_{\rm NN}}}$ = 5.02 \TeV}
\ShortTitle{Dielectron production in \pp~and \pPb~collisions at \snn = 5.02~\TeV}   

\Collaboration{ALICE Collaboration\thanks{See Appendix~\ref{app:collab} for the list of collaboration members}}
\ShortAuthor{ALICE Collaboration} 

\begin{abstract}
The first measurements of dielectron production at midrapidity (\etarangee{0.8}) in proton--proton and proton--lead collisions at \fivenn at the LHC are presented. The dielectron cross section is measured with the ALICE detector as a function of the invariant mass \mee and the pair transverse momentum \ptee, in the ranges \meeMax{3.5} and \pteeMax{8}, in both collision systems. In proton--proton~collisions, the charm and beauty cross sections, are determined at midrapidity from a fit to the data with two different event generators. This complements the existing dielectron measurements performed at \s = 7 and 13 TeV. The slope of the \s dependence of the three measurements is described by FONLL calculations. 
The dielectron cross section measured in proton--lead collisions is in agreement, within the current precision, with the expected dielectron production without any nuclear matter effects for \ee~pairs from open heavy-flavor hadron decays.
For the first time at LHC energies, the dielectron production in proton--lead and proton--proton collisions are directly compared at the same \snn via the dielectron nuclear modification factor \RpPb. The measurements are compared to model calculations including cold nuclear matter effects, or additional sources of dielectrons from thermal radiation.

\end{abstract}
\end{titlepage}

\setcounter{page}{2} 


\section{Introduction} 

ALICE~\cite{Aamodt:2008zz}, located at the Large Hadron Collider (LHC) at CERN, was designed to study  the quark--gluon plasma (QGP), a state of matter which consists of deconfined quarks and gluons. The QGP is created at the high-energy densities and temperatures reached in ultra-relativistic heavy-ion collisions. Under these conditions, the chiral symmetry is expected to be restored in the QGP phase~\cite{Karsch:1998qj,Rapp:2009yu}. Dileptons (\llepton, i.e. \ee or $\mu^{+}\mu^{-}$) are emitted during all stages of the heavy-ion collision and carry information about the medium properties at the time of their emission, as they do not interact strongly. This makes them a very promising tool to understand the chiral symmetry restoration and the thermodynamical properties of the QGP. 
In particular, the measurement of the dilepton invariant mass ($m_{\rm ll}$) allows for the separation of the different stages of the medium evolution. For \meeMax{1.1}, the main dilepton sources are Dalitz decays of pseudoscalar mesons ($\pi^{0}$, $\eta$, $\eta$') as well as Dalitz and two-body decays of vector mesons ($\rho$, $\omega$, $\phi$). In this mass range, the dilepton spectrum is sensitive to the in-medium modification of the $\rho$ meson spectral function, which is connected to the partial restoration of chiral symmetry in the hot hadronic phase~\cite{Rapp:2009yu,Tserruya:2009zt}.
At the same time, thermal radiation from the medium, contributing over a broad mass range, provides insight into the temperature of the medium and its space-time evolution. 

Measurements of dilepton production in nucleus--nucleus collisions were performed at the Super Proton Synchroton at CERN, among others, by CERES~\cite{Agakishiev:1995xb,Adamova:2006nU} and NA60~\cite{PhysRevLett.96.162302} at a \mbox{center-of-mass} energy per nucleon--nucleon pair, \snn $\leq~17.3$~GeV. An excess of dileptons compared to the expectation from known hadron decays was observed. It can be ascribed to thermal sources, primarily from thermal production of $\rho$ mesons for $m_{\rm ll} <$~1~\GeVcc with a strongly broadened $\rho$ spectral function~\cite{PhysRevLett.96.162302}, as well as partonic thermal radiation for $m_{\rm ll} >$ 1 \GeVcc~\cite{NA60QGPradiation}. At higher energies \srichnn, results from PHENIX~\cite{Adare:2015ila} and STAR~\cite{PhysRevLett.113.022301} at the Relativistic Heavy Ion Collider (RHIC) are also compatible with models involving a broadening of the $\rho$ spectral function. The study of thermal radiation from the QGP in the intermediate-mass region (IMR), \mllRange{1.1}{2.7}, is however challenging at these \mbox{center-of-mass} energies due to the large background from correlated \llepton~pairs originating from open heavy-flavor hadron decays. The first measurement of low-$m_{\rm ll}$ dileptons at the LHC, performed by ALICE in Pb--Pb collisions at \twosevensixnn~\cite{PhysRevC.99.024002}, does not provide data sensitive to a thermal signal due to the limited statistical precision and the limited knowledge of the charm contribution. Consequently, it is crucial to understand the dilepton production in proton--proton (\pp) collisions, in particular the contribution from heavy-flavor hadron decays, in order to single out the characteristic signals of the QGP.

In \pp collisions, the production of charm and beauty quarks can be estimated with perturbative quantum chromodynamics (pQCD) calculations in vacuum without any initial- and final-state effects. Owing to flavor conservation, the heavy quarks can only be produced in pairs. The resulting lepton pairs originating from charm hadron decays reflect the initial kinematic correlations between the charm and the anti-charm quarks, whereas in the case of beauty hadron decays the correlation is weakened because of their large masses. In \pp collisions, where no thermal dilepton sources are expected, the \llepton-pairs arising from heavy-flavor hadron decays are the main contribution to the dilepton yield in the IMR. Hence, dileptons can be used to study the heavy-quark production mechanisms. Together with the measurements of single heavy-flavor hadrons and their decay products, accurate results on dilepton production can provide constraints on the Monte Carlo (MC) event generators aiming to describe heavy-flavor production. In studies with dielectrons by PHENIX in \pp~collisions at \srichnn~\cite{PhysRevC.91.014907,Adare:2017caq}, and more recently by ALICE in \pp~collisions at \s = 7 and 13 TeV~\cite{pp7tev,pp13tev} the charm and beauty cross sections at midrapidity and in the full phase space were extracted by means of the analysis of the dielectron invariant mass (\mee) and pair transverse momentum (\ptee) spectra. The measured cross sections at the LHC and RHIC were found to be consistent with fixed order plus next-to-leading logarithms (FONLL) calculations~\cite{Cacciari:1998it}.

The production of dileptons in heavy-ion collisions can be modified with respect to \pp~collisions not only by the presence of hot nuclear matter but also by the presence of cold nuclear matter (CNM). The CNM effects include the modification of the quark and gluon content in the initial state, that is described by means of parton distribution functions (PDFs) of the incoming nucleons in the collinear factorization framework. In nucleons that are bound in the nucleus, the PDFs are altered by the presence of additional nuclear matter with respect to free nucleons. This modification depends on the parton momentum fraction $x$, the atomic mass number of the nucleus $A$, and the momentum transfer $Q^{\rm 2}$ in the hard scattering process. 
Nuclear PDFs are obtained from a global fit to data from different experiments~\cite{Hirai:2007sx,EPS09,Eskola:2016oht}. When the phase space density of gluons within the hadron is high due to gluon self-interactions, reaching a saturation regime, an appropriate theoretical description is the Color Glass Condensate (CGC) theory~\cite{Gelis:2010nm,Tribedy:2011aa,Albacete:2012xq,Rezaeian:2012ye}. 
At LHC energies at midrapidity, where small values of $x$ are probed by the charm and beauty production ($x \leq 10^{-3}$), the most relevant effect on the PDFs is shadowing~\cite{nucShadowing}. The modification of the initial state in hadronic collisions can significantly reduce the heavy-flavor production cross sections at low transverse momentum (\pt). In addition, multiple scattering of partons in the nucleus, before and/or after the hard scattering, can change the kinematic distribution of the produced hadrons and affect their azimuthal correlation, such that the \mee~and \ptee distributions from correlated heavy-flavor hadron decays could be modified~\cite{Vitev:2007ve,Kopeliovich:2002yh}.

Initially, hot matter effects were not expected in proton--nucleus (\pA) collisions, so they were used as a baseline for measurements in heavy-ion collisions to study possible CNM effects. At LHC energies in minimum bias (MB) \pPb~collisions at midrapidity, the measured \pt differential production cross sections of single open-charm hadrons~\cite{Dmeson2016,Dmeson2019} and their decay electrons~\cite{Adam:2015qda,HFelec2020}, as well as results on azimuthal correlations of D mesons and charged particles~\cite{ALICE:2016clc}, are compatible over the whole \pt range probed with the results in \pp collisions scaled with the atomic mass number $A$ of the Pb nucleus. Moreover, the yields of \jpsi from B hadron decays as well as prompt \jpsi are found to be suppressed at low \pt at midrapidity in MB \pPb~collisions at \fivenn~\cite{Acharya:2018yud}, but the measurements of B hadron production cross sections at high \pt show no significant modification of the spectra compared to perturbative QCD calculations of \pp collisions scaled with $A$. All of these results indicate that possible CNM effects are small compared to the current uncertainties of the measurements for open heavy-flavor production at midrapidity at the LHC. However, at forward and backward rapidities, the measured \pt differential cross sections of D~\cite{Aaij_2017} and B mesons~\cite{Aaij:2017cqq}, and of muons originating from heavy-flavor hadron decays~\cite{Acharya:2017hdv} in minimum bias \pPb~collisions at \fivenn demonstrate the presence of CNM effects and support shadowing as possible explanation. The forward and backward results set constraints on models that also aim at reproducing the midrapidity measurements. Accurate measurements in \pA collisions provide important inputs for the parametrizations of the nuclear PDFs, which are currently suffering from large uncertainties~\cite{EPS09,Eskola:2016oht}.

On the other hand, final-state effects may also play an important role in \pA collisions. In particular, in those with large multiplicities of produced particles, as suggested by results from azimuthal anisotropy measurements through two-particle~\cite{CMS:2012qk,Abelev:2012ola,ABELEV:2013wsa,Aad:2012gla,Adam:2015bka,Sirunyan:2018toe,Acharya:2018dxy} and multi-particle correlations~\cite{Khachatryan:2015waa,Aaboud:2017blb}, modifications of the \pt distributions of identified hadrons with respect to the charged-particle multiplicity in the event~\cite{Abelev:2013haa,Chatrchyan:2013eya}, multiplicity dependence of strangeness production~\cite{strangeenhancement}, and $\psi$(2S) production~\cite{Abelev:2014zpa,Aaij:2016eyl,Adam:2016ohd}. Should such observations be linked to the creation of a small volume of hot medium in high-multiplicity \pA collisions, the corresponding thermal radiation could lead to an enhanced dilepton production~\cite{Rapp:2000pe,Rapp:2013nxa,Shen:2016zpp}.
At RHIC energies, results on dilepton production at midrapidity in minimum bias d--Au collisions at \snn = 200 \GeV~\cite{PhysRevC.91.014907,Adare:2017caq}  show no evidence of neither an additional source of lepton pairs, nor of nuclear modification of the charm and beauty production. At the LHC, where the density of final-state particles is larger, dilepton measurements in \pPb collisions can give more insight into the possible formation of a hot medium in small systems and CNM effects.

In this article, the first measurements of \ee production in \pp and \pPb collisions at \fivenn at the LHC are presented. The results are obtained with the ALICE detector. The data are compared, in terms of the \mee and \ptee distributions, to the sum of the expected sources of \ee~pairs from known hadron decays, the so-called hadronic cocktail. 
The spectra are shown after the application of fiducial requirements on single electrons (\etarangee{0.8} and \pteRange{0.2}{10}) without an extrapolation to the full phase space. In addition, for the first time at LHC energies, a direct comparison between the dielectron cross section obtained in \pp~and \pPb~collisions is possible since both data sets were recorded at the same \snn. 
In particular, the analysis of the \pp data resolves the model dependence on the expected \mee and \ptee distributions of correlated \ee~pairs from open heavy-flavor hadron decays in \pp~collisions, used as reference for the \pPb~study. This allows for the research of possible modifications to the dielectron production in \pPb~collisions due to CNM or additional final-state effects.

The article is organized as follows. The experimental setup and the used data samples are described in Sec.~\ref{chap:detData}. The analysis steps, including track selection criteria, electron identification, signal extraction and efficiency corrections, are described in Sec.~\ref{chap:analysis}, together with the corresponding systematic uncertainties. The method to calculate the expected dielectron cross section from known hadron decays is explained in Sec.~\ref{cocktail}. In Sec.~\ref{chap:results}, the results are presented, covering the charm and beauty cross section extracted in \pp~collisions, comparisons of the dielectron production in \pp~and \pPb~collisions to the expectations from known hadron decays, and the resulting dielectron nuclear modification factors.

\section{The ALICE detector and data samples}
\label{chap:detData}

The ALICE detector and its performance are described in~\cite{Aamodt:2008zz,Abelev:2014ffa}. Electrons are measured in the ALICE central barrel covering the midrapidity range $|\eta| < 0.9$. (Note that the term `electron' is used for both electrons and positrons throughout this paper.) The relevant subsystems used in the dielectron analysis are the Inner Tracking System~(ITS)~\cite{Aamodt:2010aa}, the Time Projection Chamber~(TPC)~\cite{ALME2010316}, and the Time-Of-Flight~(TOF)~\cite{Akindinov:2013tea} detector.

The innermost detector of the ALICE apparatus, closest to the nominal interaction point, is the ITS. It consists of six silicon tracking layers based on three different technologies. The two inner layers are Silicon Pixel Detectors (SPD), the two middle layers are Silicon Drift Detectors, and the two outer layers are Silicon Strip Detectors. About half of the \pp~and \pPb~data samples were recorded without the Silicon Drift Detector information in order to reach maximal data acquisition rates. For this reason, even when available, the information from this detector is not used so as to have uniform detector conditions over the entire data sets. 
The main detector for particle identification (PID) and tracking is the TPC. This 500 cm long cylindrical detector, with an outer radius of 247 cm, is located around the ITS. The TPC readout is based on multi-wire proportional chambers and provides up to 159 three-dimensional space points as well as the specific energy loss of the particle.
The outermost detector used in this analysis is the TOF. It provides a time-of-flight measurement for particles from the interaction point to its active volume, at a radius of 370~cm.
The combined information from the ITS, TPC, and TOF is used to reconstruct the track of a charged particle using a Kalman-filter based algorithm~\cite{Abelev:2014ffa}.

The data used in this paper were recorded in collisions at \fivenn, with the \pPb~data taken in 2016, and the pp data taken in 2017. 
Due to the asymmetric beam energies in the \pPb~configuration, 4~TeV for the proton beam and 1.59~TeV per nucleon for the Pb beam, the rapidity ($y$) of the \mbox{center-of-mass} system is shifted by $\Delta y = 0.465$ in the laboratory frame in the direction of the proton beam. For both collision systems, events were recorded when a coincident signal in the \VZERO detector system~\cite{Abbas:2013taa} was registered. The \VZERO detector consists of two segmented scintillators located at $+340$~cm and $-70$~cm along the beam axis from the nominal interaction point.
Additional selections are applied to the recorded events. The background from beam--gas interactions and pileup events are rejected by using the correlations between the \VZERO detector and ITS signals. Only events with at least one track segment reconstructed in the ITS contributing to the vertex reconstruction with the SPD are used. To assure a uniform detector coverage at midrapidity, the vertex position along the beam direction is restricted to $\pm10$~cm with respect to the nominal interaction point. A summary of the number of events $N_{\rm ev}$ passing the event selection criteria and the corresponding integrated luminosity $\lumi_{\rm int}$ is given in Table~\ref{tab:datasets}. These requirements are fulfilled by 77\% (75\%) of the recorded events for the \pp~(\pPb) data samples. 
The $\lumi_{\rm int}$ is calculated as $\lumi_{\rm int}$ = $N_{\rm MB}$ $/$ $\sigma_{\rm MB}$, with the number of analyzed events after the vertex reconstruction efficiency correction $N_{\rm MB}$, and the minimum bias trigger cross section $\sigma_{\rm MB}$ measured via a van der Meer scan in the corresponding collision system~\cite{ALICE-PUBLIC-2016-005,Abelev:2014epa}.

\begin{table}[htb!]
    \centering
    \caption{The integrated luminosity ($\lumi_{\rm int}$) and the number of events ($N_{\rm ev}$) after event selection criteria are applied for the \pp~and \pPb~data samples.}
    \renewcommand{\arraystretch}{1.3}
    \begin{tabular}{@{\extracolsep{3mm}} c c c c }
    \hline
    \hline
        Data set & $\lumi_{\rm int}$ & $N_{\rm ev}$ \\
        \hline
        \pp~& 19.93$\pm$0.4~$\rm{nb^{-1}}$& 888$\times$10$^{6}$\\
        
        \pPb~&  299$\pm$11~$\rm{\mu b^{-1}}$& 535$\times$10$^{6}$\\
        \hline
        \hline
    \end{tabular}
    \label{tab:datasets}
\end{table}

\section{Data analysis}
\label{chap:analysis}
\subsection{Track selection}
The same track selection criteria are applied in the analysis of the \pp~and \pPb~data samples. Electron candidates are selected from charged tracks reconstructed in the ITS and TPC in the transverse-momentum range \pteRange{0.2}{10} and pseudorapidity range \etarangee{0.8}.
The tracks are required to have at least 80 space points reconstructed in the TPC and at least three hits in the ITS assigned to them. The maximum $\chi^2$ per space point measured in the TPC (ITS) is required to be smaller than 4 (4.5). 
To reduce the contribution from secondary tracks, the distance-of-closest approach of the track to the reconstructed primary vertex is required to be smaller than 1~cm in the transverse plane to the colliding beams and smaller than 3~cm in the longitudinal direction. In order to further suppress the contribution of electrons from photon conversions in the detector material, only tracks with a hit in the first layer of the SPD and no ITS cluster shared with any other reconstructed track are used in the analysis.

\subsection{Electron identification}
Electrons are identified by measuring their specific energy loss \dEdx in the \TPC and their velocity with the \TOF as a function of their momentum. The momentum is estimated from the curvature of the track measured in the ITS and TPC.
The PID is based on the detector PID response $n (\sigma_{i}^{\rm Det})$. This is expressed as the deviation between the measured PID signal of the track in the detector (Det) and its expected most probable value for a given particle hypothesis $i$ at the measured track momentum. This deviation is normalized to the detector resolution $\sigma$.  
Electrons are selected over the whole investigated momentum range in the interval $|n (\sigma_{e}^{\rm TPC})| < 3$, while
 the charged pion ($\pi^{\pm}$) contribution is suppressed by requiring $n (\sigma_{\pi}^{\rm TPC}) > 3.5$. Furthermore, the track must also fulfill at least one of the two following conditions:
\begin{enumerate}[]
\vspace{-0.4cm}
\item The track is outside the hadron bands in the TPC, defined by $|n (\sigma_{\rm K}^{\rm TPC})| < 3$ and\\ ${|n (\sigma_{\rm p}^{\rm TPC})| < 3}$.
\vspace{-0.2cm}
\item The track has a valid hit in the TOF detector and falls within the range  $|n (\sigma_{e}^{\rm TOF})| < 3$.
\vspace{-0.4cm}
\end{enumerate}
With this approach, the hadron contamination in the single-electron candidate sample is less than 4\% averaged over \pte. The largest hadron contamination, up to 9\%, is observed where kaons ($\pt \approx$~0.5~\GeVc), protons ($\pt \approx $~1~\GeVc) or charged pions (\ptMin{7}) have a similar \dEdx as electrons in the TPC. The final hadron contamination in the dielectron signal is negligible, as pairs containing a misidentified hadron are further removed during the signal extraction.

\subsection{Signal extraction}
A statistical approach is used to extract the true signal pairs ($S$) as a function of \mee and \ptee, in which all electrons and positrons in an event are combined to create an opposite-sign spectrum ($OS$). The $OS$ contains not only signal, but also background ($B$) from combinatorial pairs, as well as residual correlations from jets and conversions of correlated decay photons originating from the same particle. The background is estimated from the distribution of same-sign pairs ($SS$) from the same event, as explained in~\cite{pp7tev}.
The advantages of the same-sign technique, with respect to an event-mixing approach, are the intrinsic correct normalization of the $SS$ spectrum, and the inclusion of charge-symmetric background sources, e.g. electrons from fragmentation in jets. The signal is then extracted as $ S = OS - R_{\rm acc} \cdot SS$, where $R_{\rm acc}$ is a correction factor needed to account for the different acceptance of opposite-sign and same-sign pairs. It is estimated using an event-mixing technique detailed in~\cite{pp7tev}.

For pairs with \meeMax{0.14}, the angle \phiV, which quantifies the orientation of the opening angle of the pairs relative to the magnetic field~\cite{pp7tev} and allows for the rejection of \ee~pairs from photon conversions, is required to be smaller than 2 rad. After applying this criterion, the remaining contribution from \ee~pairs from photon conversions in the detector material is less than 1.4\%.

\begin{figure}[!t]
    \begin{center}
    \includegraphics[width=0.49\textwidth]{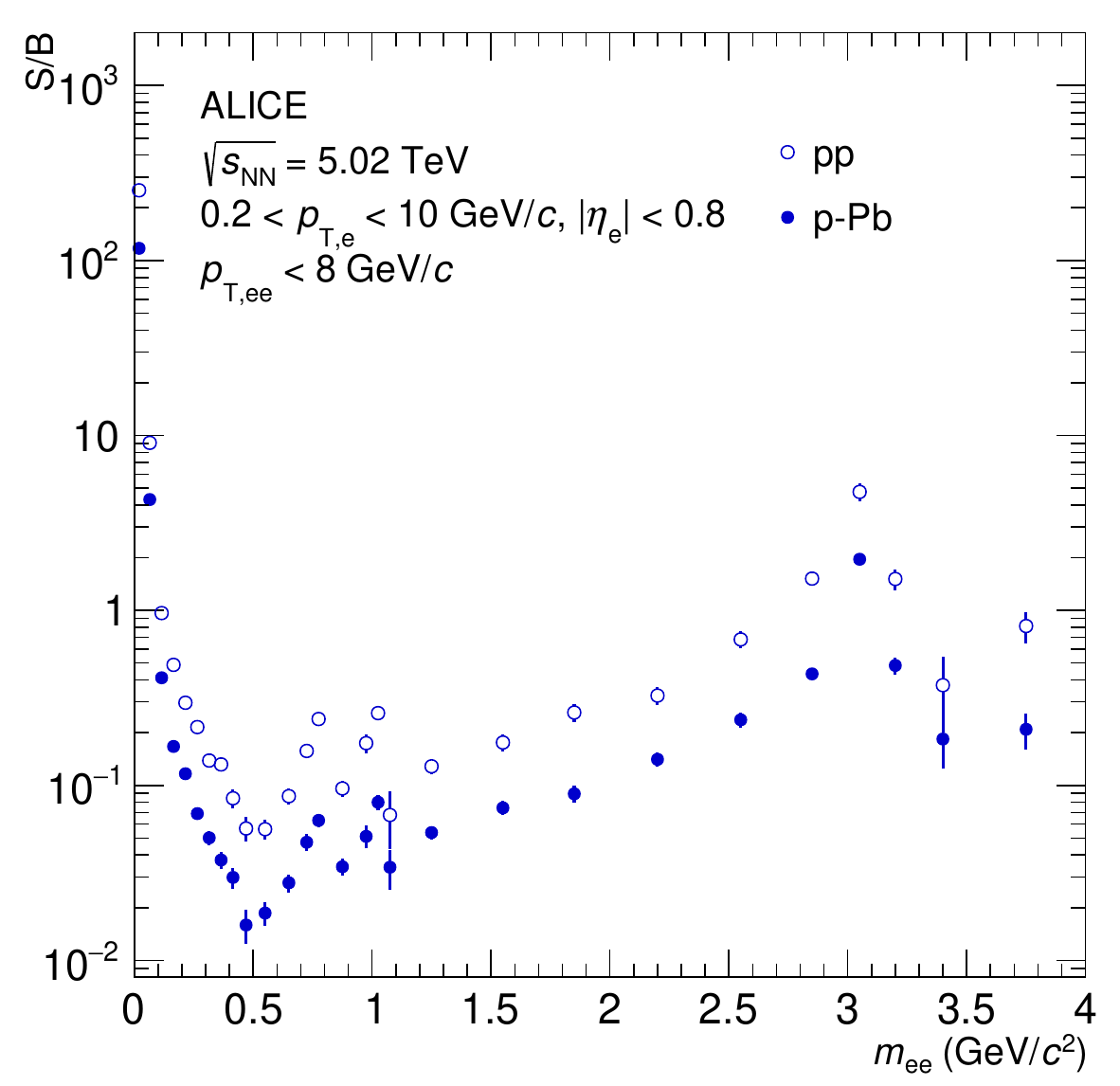}
    \includegraphics[width=0.49\textwidth]{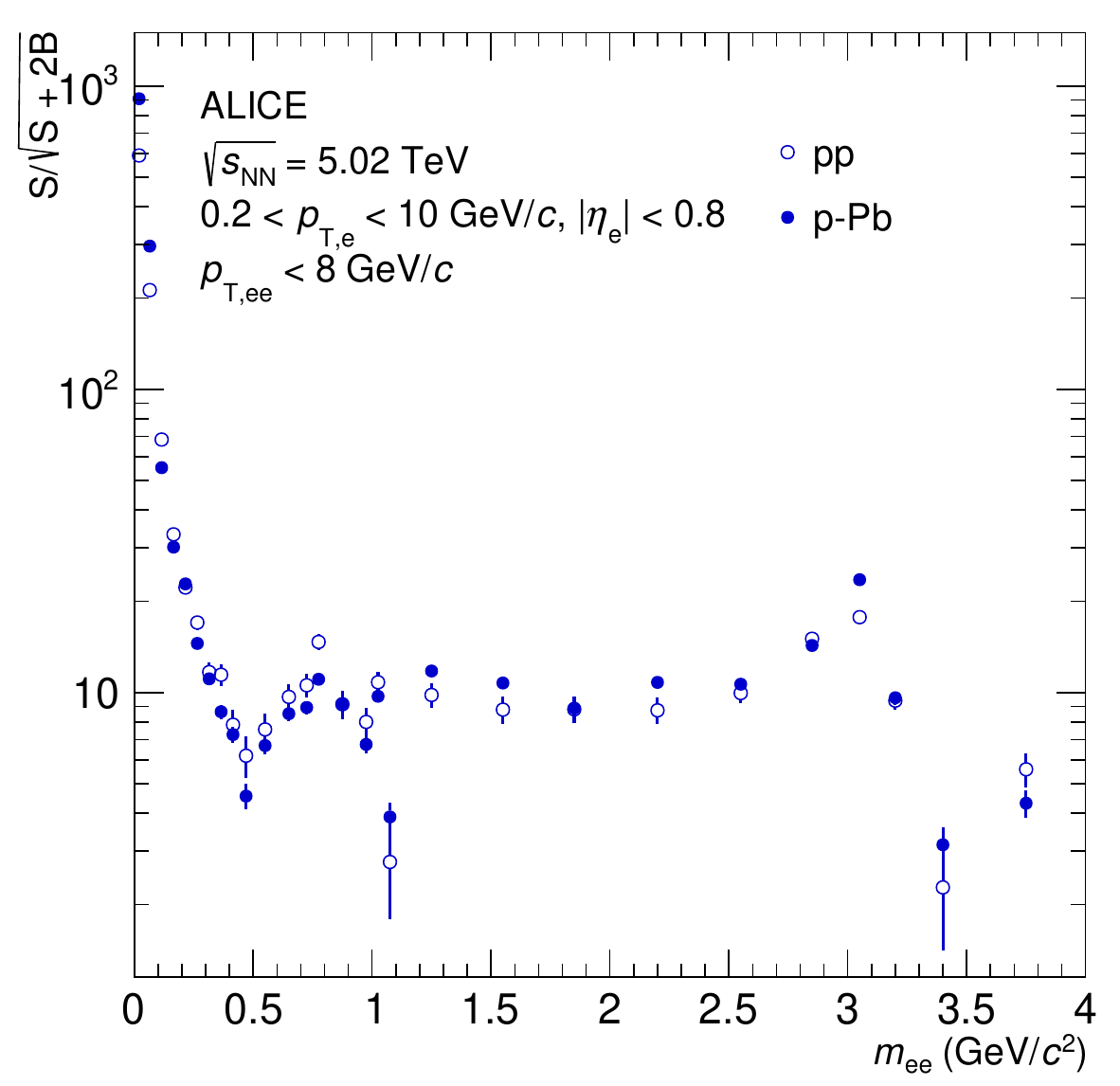}
    \end{center}
    \caption{(Color online) Signal-to-background ratio (left) and statistical significance (right) of the dielectron measurements as a function of \mee in \pp~and in \pPb~collisions at \fivenn.}
    \label{fig:soverb}
\end{figure}

The signal-to-background ratio ($S/B$) and statistical significance ($S/ \sqrt{S + 2B}$) are depicted in the left and right panels of Fig.~\ref{fig:soverb}, respectively, for the \pp~and \pPb~samples. Despite a worse $S$/$B$ in \pPb~collisions, mostly due to the larger particle multiplicity, the statistical significance of the measurement is similar in both collision systems. 

\subsection{Efficiency correction}

The efficiency of the single-electron and pair selection is calculated with dedicated MC simulations. The simulated events are propagated through the ALICE detector using the GEANT~3~\cite{Brun:118715,Brun:1994aa} transport code. 
The same strategy is used for the \pp~and \pPb~analyses. Since the full kinematic range cannot be fully populated by pairs originating only from the same-mother particle (SM) or only from open heavy-flavor hadron decays (HF), the final efficiency correction is estimated separately for each source.
For SM pairs, \pp~and \pPb~collisions are generated with the Monash2013~\cite{monash2013} tune of PYTHIA~8.1~\cite{pythia81} (denoted as PYTHIA 8 from now on) and with DPMJET~\cite{dpmjet}, respectively. In the case of HF pairs, MC simulations of open heavy-flavor hadrons using PYTHIA 6~\cite{pythia64} are performed. 
In the \pPb~case, heavy-flavor events are embedded into realistic \pPb~collisions simulated with EPOS-LHC~\cite{eposlhc}. The efficiency as a function of \mee and \ptee is calculated as
\begin{equation}
    \epsilon_{\rm ee}(\mee,\ptee) = w_{\rm SM} \cdot \epsilon_{\rm SM \rightarrow ee}(\mee,\ptee) + w_{\rm HF} \cdot \epsilon_{\rm HF \rightarrow ee}(\mee,\ptee).
\end{equation}
The weights $w_{\rm SM}$ and $w_{\rm HF}$ represent the relative cross sections of the SM and HF sources, respectively. They are estimated with the expected dielectron cross section from known hadron decays, explained in Sec.~\ref{cocktail}. The average reconstruction efficiency of a signal \ee~pair is very similar in \pp~and \pPb~collisions and ranges from 25\% to 30\%.

The corrected differential dielectron cross section is calculated as
\begin{equation}
    \frac{{\rm d}^{2}\sigma_{\rm ee}}{{\rm d}\mee {\rm d}\ptee} = \frac{1}{\mathcal{L}_{int}}\frac{1}{\Delta \mee}\frac{1}{\Delta \ptee} \frac{S(\mee,\ptee)}{\epsilon_{\rm ee}(\mee,\ptee)},
\end{equation}
with $\Delta \mee$ and $\Delta \ptee$ being the width of the \mee and \ptee intervals, respectively, and $\mathcal{L}_{int}$ the integrated luminosity. In \pp~collisions, the spectra are corrected for the vertex reconstruction efficiency and for the efficiency of the minimum bias trigger to select inelastic events with an \ee pair, which are found to be 96\% and 98\%, respectively. In \pPb~collisions both efficiencies are unity.

\subsection{Systematic uncertainties}

Different sources of systematic uncertainties are taken into account. On the single track level the effects of the required hit in the first ITS layer, the ITS-TPC matching efficiency, and the selection of tracks without shared clusters are studied. These uncertainties are calculated as a function of \mee~and \ptee. Effects from the track and PID selection as well as the requirement on \phiV~are estimated on the pair level. For these uncertainties negligible \ptee~dependence is found and they are applied only as a function of \mee.
In order to suppress statistical fluctuations, that could influence the estimated systematic uncertainties, they are evaluated in both analyses in wide mass intervals.
The resulting systematic uncertainties from the different sources are summarized in Table~\ref{tab:summarySyst} for the \pPb~and \pp~analyses.
 
The systematic uncertainties that arise from the limited knowledge of the matching efficiency of the track segments reconstructed in the ITS and the TPC, and from the requirement of a hit in the innermost ITS layer, are determined with a two-step procedure. First, on the single-track level, the efficiencies of these two track selection criteria are estimated for charged pions in data and in MC as a function of \pt. Second, the observed difference is taken as input to a toy MC simulation, which generates particles in the full \mee and \ptee phase space decaying them into \ee~pairs and applying the fiducial selection. The final systematic uncertainty at the pair level is then calculated as the sum of the uncertainties of the decay products, corresponding to the input.   

The systematic uncertainty from the requirement of no shared clusters in the ITS is evaluated by varying the maximum number of allowed shared ITS clusters for the selected electron candidates. This provides a test of the understanding of the background since it not only probes different single-electron efficiencies but also different $S/B$ ratios. When no requirement is applied the $S/B$ decreases by a factor two, which is due to the increased contribution of electrons from photon conversions in the detector material in the selected electron sample. The resulting dielectron spectra are compared after the efficiency correction.
The maximum deviation of the variations that are considered statistically significant according to the Barlow criterion~\cite{BARLOW1993219} is used to assign the systematic uncertainty. 

Similarly, the uncertainty from the remaining single-electron selection criteria is determined by varying them simultaneously within reasonable values.
By changing the selection criteria for the tracks in the ITS and the hadron rejection criteria, the evaluated systematic uncertainties are sensitive to estimations of the background as well as a possible bias due to the hadron contamination in the electron sample.
The systematic uncertainty is calculated as the root mean square of the variation of the final data points. Finally, a possible bias due to the efficiency correction of the \phiV~selection is estimated. For this purpose, the maximum \phiV~requirement for \ee~pairs with \meeMax{0.14} is varied around its default value from 1.5 to 2.7 rad.

Two additional sources of uncertainty are taken into account for the \pp~analysis, namely the correction for the primary vertex reconstruction efficiency and the trigger efficiency. Both are evaluated to be 2\% based on MC simulations. A priori, the reconstruction efficiency of an \ee~pair at a given \mee and \ptee should not depend on its source. However, in the \pPb~analysis, a difference in the efficiencies of \ee~pairs originating from either light-flavor decays or heavy-flavor decays is observed. Therefore, an additional uncertainty of 3\% is assigned to cover a possible bias in the spectra. The total systematic uncertainty is calculated as the quadratic sum of the individual contributions assuming they are all uncorrelated. The total uncertainty varies between 11\% and 4\%, being equal to 5\% in most of the \mee~range. The uncertainties are partially correlated between different \mee~intervals. 
\begin{table}[htb]
    \centering
    \caption{Systematic uncertainties on the requirement of a hit in the first ITS layer, the ITS-TPC matching efficiency (ME), the allowed number of shared clusters in the ITS, the variation of the \phiV~selection, and the tracking and PID variations in coarse \mee~intervals for the \pPb(\pp) analysis. The uncertainties on the vertex reconstruction (2\%) and trigger (2\%) efficiencies in the \pp analysis, as well as the uncertainty of the light- and heavy-flavor efficiency differences (3\%) in the \pPb analysis, are not listed. They are applied over the whole range of the measurement and included in the total uncertainty.
    The total systematic uncertainty is the quadratic sum of the single contributions assuming they are all uncorrelated.}
    \renewcommand{\arraystretch}{1.3}
    \begin{tabular}{@{\extracolsep{1mm}}c c c c c c c c}
    \hline
    \hline
        \mee (\GeVcc) & 1st ITS layer & ITS-TPC ME & Shared ITS cls. & $\varphi_{\rm V}$ & Tracking \& PID & Total\\
         \hline
       ~~~~~~~ $< 0.14$  & 2 (1)\% & 2 (2)\% & 2 (1)\% & 2 (1)\% & 10 (6)\% & 11 (7)\%\\
        $0.14 - 1.1$     & 2 (1)\% & 2 (2)\% & 2 (0)\% & --     & 2 (2)\%  & 5 (4)\%\\
        $1.1 - 2.7$      & 2 (2)\% & 2 (3)\% & 0 (0)\% & --     & 2 (2)\%  & 5 (5)\%\\
        $2.7 - 3.5$      & 2 (2)\% & 2 (4)\% & 0 (0)\% & --     & 2 (1)\%  & 5 (5)\%\\
        \hline
        \hline
    \end{tabular}
    \label{tab:summarySyst}
\end{table}

\section{Cocktail of known hadron decays}
\label{cocktail}

The measured dielectron spectra in \pp~and \pPb~collisions are compared to a hadronic cocktail, which represents the sum of the expected contributions of dielectrons from known hadron decays, after the fiducial selection criteria on single electrons are employed. A fast MC simulation of the ALICE central barrel is performed, including realistic momentum and angular resolutions as well as Bremsstrahlung effects, which are applied to the decay electrons as a function of \pte, azimuthal angle ($\varphi_{\rm e}$) and $\eta_{\rm e}$~\cite{CocktailSmearing}.

\begin{figure}
 \begin{center}
    \includegraphics[scale = 0.39]{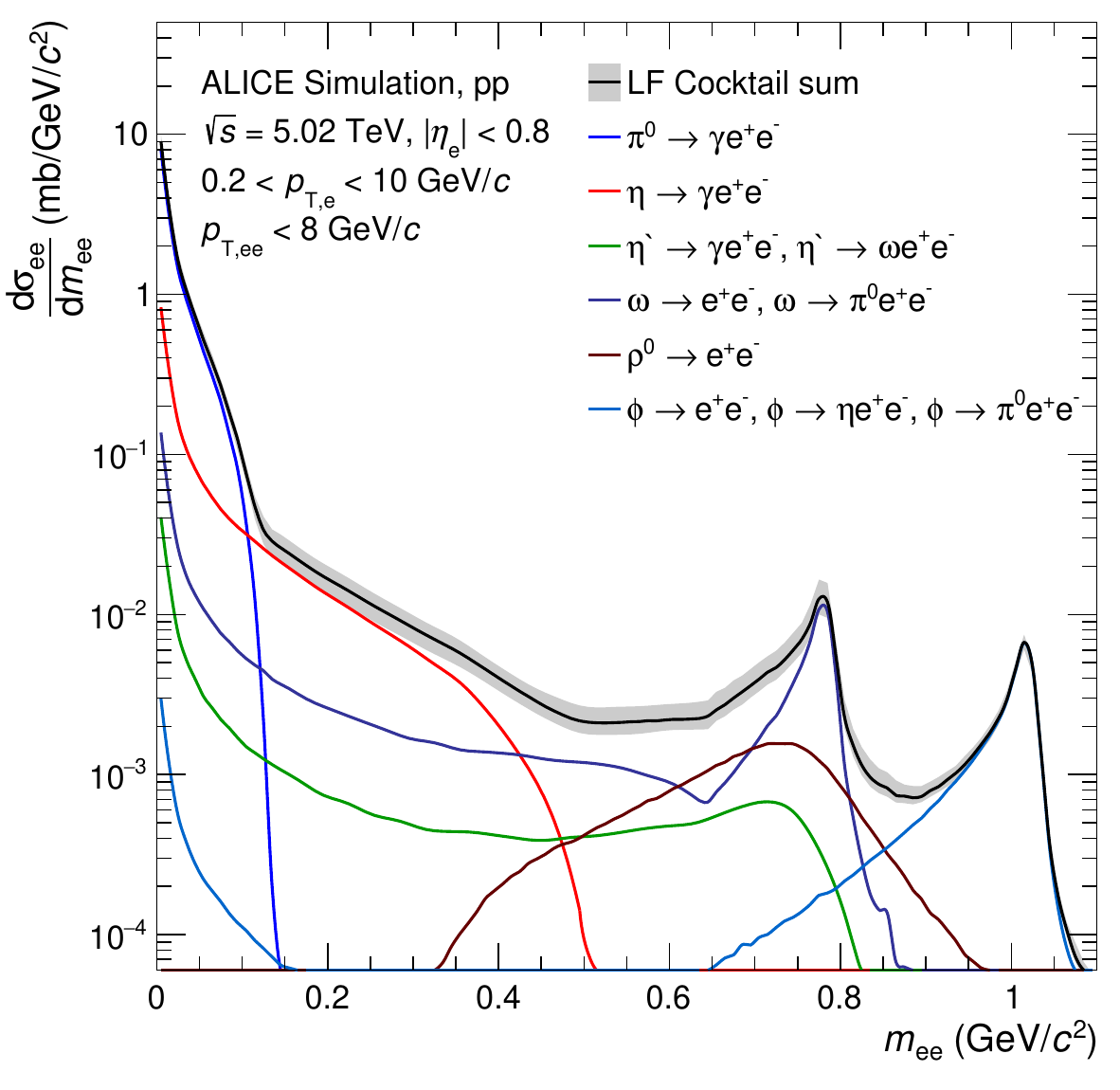}
    \end{center}
    \caption{(Color online) Expected cross section for dielectron production from light-flavor hadron decays in \pp~collisions at \fivenn as a function of \mee. The sum of the single light-flavor (LF) contributions is shown (solid black line) with its uncertainties (gray band).}
    \label{fig:lfCocktail}
\end{figure}

The Dalitz and dielectron decays of light neutral mesons are simulated with the phenomenological event generator EXODUS~\cite{phenixppAuAu}, following the approach described in~\cite{pp7tev}. The \pt spectra of light neutral mesons measured at midrapidity in \pp~collisions at different \mbox{center-of-mass} energies and in \pPb~collisions at \fivenn are parametrized and taken as input to the calculations. Since the measured \pt distributions of $\pi^{\pm}$ mesons extend to lower \pt, they are used to determine the $\pi^{0}$ input parametrizations. The \pt spectra of $\pi^{\pm}$ mesons measured by ALICE in \pp~and \pPb~collisions at \fivenn~\cite{pppi,pPbpi} are first parametrized with a modified Hagedorn function~\cite{modifiedHagedorn}. A \pt-dependent scaling factor is then applied to the $\pi^{\pm}$ parametrization in order to account for the difference between $\pi^{0}$ and $\pi^{\pm}$ due to isospin-violating decays, mainly of the $\eta$ mesons. This factor is estimated using an effective model that describes measured hadron spectra at low \pt and includes strong and electromagnetic decays. 
The measured \pt spectra of $\phi$ mesons in \pp~and \pPb~collisions at \fivenn~\cite{ppphi,pPbphi} are fitted to obtain the $\phi$ input parametrizations.  The \pt spectra of the other light mesons, $\eta$, $\eta'$, $\rho$, and $\omega$ are derived from the $\pi^{\pm}$ spectrum. 
The \pt spectrum of the $\eta$ meson is estimated from a common fit to the ratios of the $\eta$ to $\pi^{0}$ \pt spectra in \pp~collisions at \s = 7 TeV~\cite{ppeta2pion}, 8 TeV~\cite{Acharya:2017tlv}, and in \pPb~collisions at \fivenn~\cite{neutralMeson_pPb} measured by ALICE as well measurements by CERES/TAPS in p--Au and p--Be collisions at $\sqrt{s_{\rm NN}}$ = 29.1 GeV which extend to lower \pt (\ptMax{2}) ~\cite{cerestapshelios}. The \pt distributions of $\omega$ and $\rho$ are obtained from the respective ratios to the $\pi^{\pm}$ \pt distributions in simulated pp collisions at \five with PYTHIA 8. The $\eta$/$\pi^{0}$, $\rho$/$\pi^{\pm}$ and $\omega$/$\pi^{\pm}$ ratios as a function of \pt are assumed to be independent of the \pA or \pp collision system and of the energy, as suggested by the measurements~\cite{ppeta2pion,Acharya:2017tlv,neutralMeson_pPb,cerestapshelios}. Therefore, common parametrizations of these ratios are used for the \pp~and \pPb~cocktails. Finally, the $\eta'$ meson is generated assuming $m_{\rm T}$ scaling~\cite{mt1,mt2,mt3}, implying that the spectra of all light mesons as a function of $m_{\rm T} = \sqrt{m_{\rm 0}^{2} + \pt^{2}}$, where $m_{\rm 0}$ is the pole mass of the considered mesons, follow the same shape and only differ by a normalization factor.
All contributions from the decays of light-flavor hadrons as a function of \mee are shown in Fig.~\ref{fig:lfCocktail}. In order to estimate the \jpsi contribution, the measured \jpsi \pt spectra in \pp~and \pPb~collisions at \fivenn~\cite{ppjpsi,pPbjpsi} are parametrized and used as inputs for the simulations. The \jpsi mesons are decayed using PHOTOS~\cite{photos} via the dielectron channel, which also includes the full QED radiative channels. 

{}
The contributions of correlated semileptonic decays of open charm and beauty hadrons are calculated with two different MC event generators. They are identical to the ones used in the dielectron analyses performed by ALICE in \pp collisions at $\sqrt{s}$ = 7 TeV~\cite{pp7tev} and $\sqrt{s}$ = 13 TeV~\cite{pp13tev}: PYTHIA~6.4~\cite{pythia64} with the Perugia2011 tune~\cite{perugia2011} and the next-to-leading order event generator POWHEG~\cite{powheg1,powheg2,powheg3,powheg4} with PYTHIA 6 to evolve the parton shower. Only the shapes of the expected \mee and \ptee dielectron spectra are estimated with the MC event generators. The absolute normalization is obtained from a fit of the measured dielectron cross sections in pp collisions at \five, as shown in Sec.~\ref{resultsPP}. For \pPb~collisions, the \ccbar and \bbbar cross sections, extracted in \pp~collisions, are scaled with the atomic mass number $A$ of the Pb nucleus (208). This approach neglects any cold nuclear matter effects, which will be discussed in Sec.~\ref{resultsPPB}. 

The following sources of systematic uncertainties are taken into account: the input parametrizations of the measured $\pi^{\pm}$, $\phi$ and \jpsi \pt spectra and $\eta/\pi^{0}$ ratios, the scaling factor applied to the $\pi^{\pm}$ parametrizations, the $m_{\rm T}$ scaling parameters, and the different decay branching ratios. The uncertainty of the $\pi^{\pm}$ scaling factor is estimated from variations of the model parameters.
For the $\rho$ and $\omega$ mesons, the uncertainty of the $\omega/\pi^{0}$ and $\rho/\pi^{0}$ ratios are estimated by comparing the measured and simulated ratios in \pp collisions at $\sqrt{s}$ = 7 TeV~\cite{ALICE-PUBLIC-2018-004} and $\sqrt{s}$ = 2.76 TeV~\cite{Acharya:2018qnp}, respectively. The total uncertainties of the \pp~and \pPb~cocktails vary from 5\% to 20\% depending on the \mee and \ptee interval.

\section{Results}
\label{chap:results}

The dielectron cross sections in \pp and \pPb~collisions as well as the nuclear modification factor are presented differentially as a function of \mee for \pteeMax{8} and as a function of \ptee in two different mass regions, the low-mass region (LMR), \meeRange{0.5}{1.1}, and the intermediate-mass region (IMR), \meeRange{1.1}{2.7}.

\subsection{Heavy-flavor cross sections in \pp~collisions}
\label{resultsPP}

\begin{figure}[tb]
    \begin{center}
    \includegraphics
    [width = 0.49\textwidth]{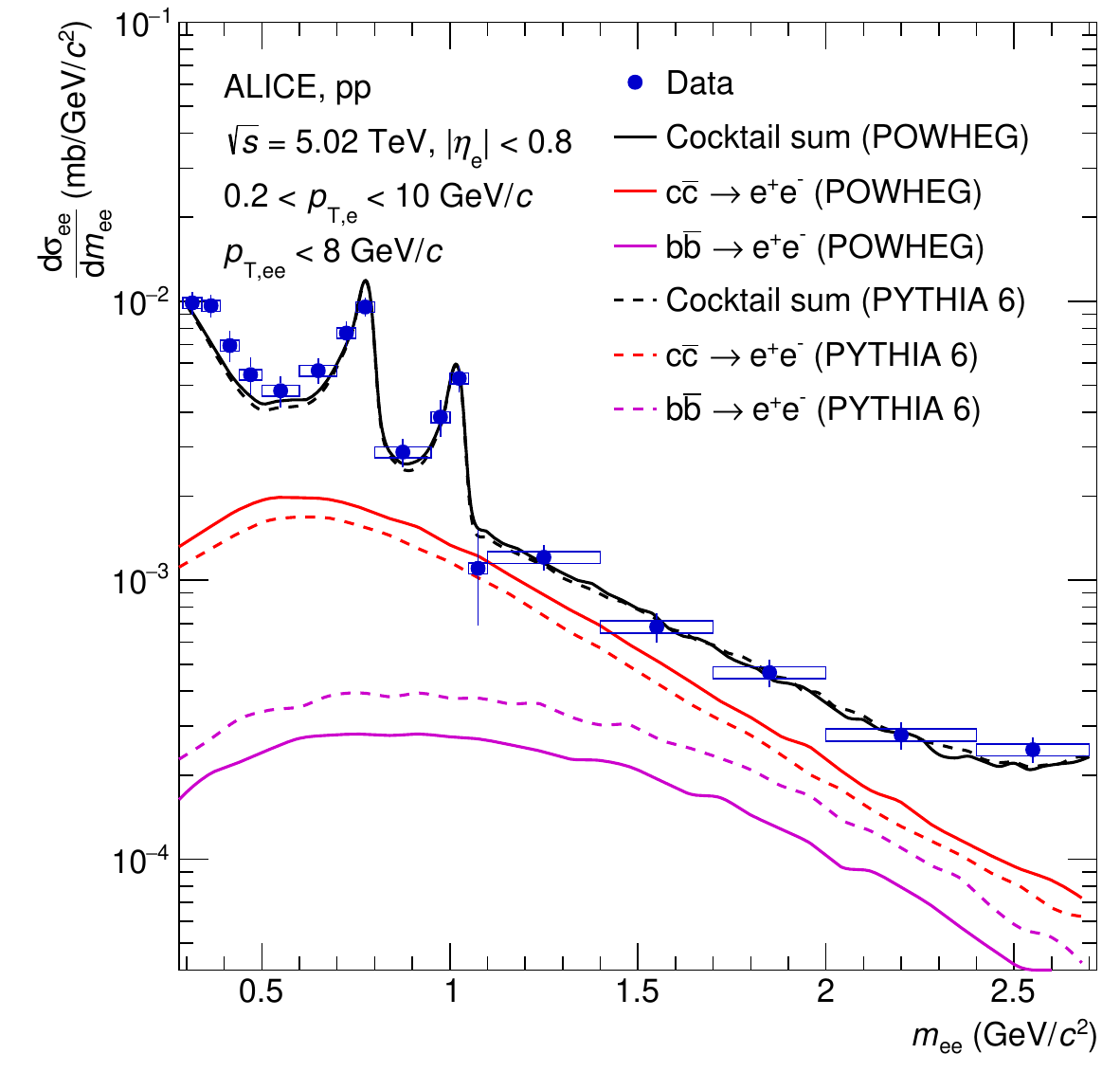}
    \includegraphics
    [width = 0.49\textwidth]{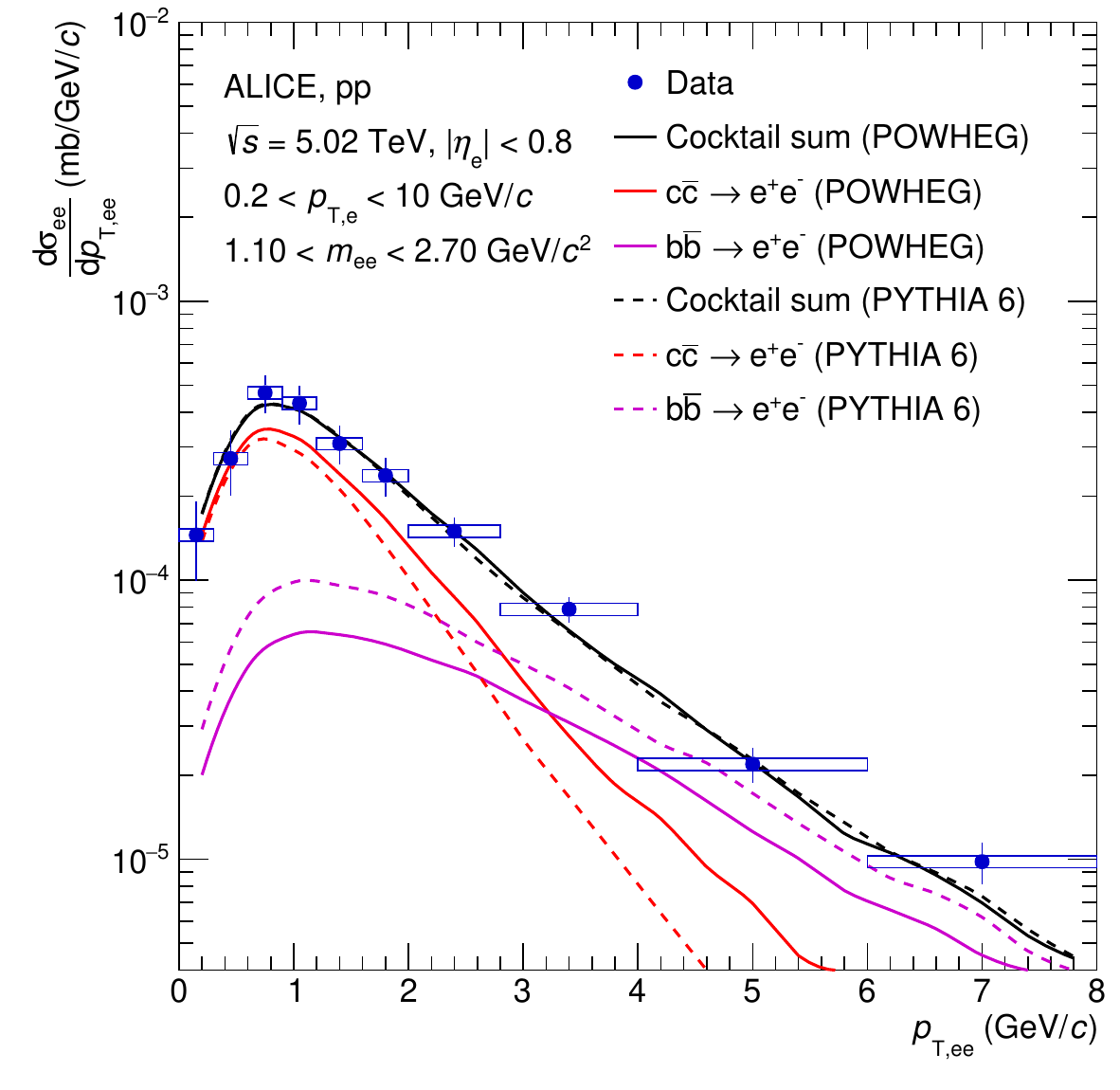}
    \end{center}
    \caption{(Color online) Projections of the heavy-flavor dielectron fits as a function of \mee (left) and \ptee (right) using POWHEG (solid black line) and PYTHIA 6 (dashed gray line) as event generators. The colored lines show the charm (red) and beauty (magenta) contributions for both event generators after the fit.}
    \label{fig:fits_pps}
\end{figure}

The differential \ee~production cross sections d$\sigma_{\rm ee}$/d$m_{\rm ee}$ and d$\sigma_{\rm ee}$/d$p_{\rm T,ee}$ in pp collisions, measured in the IMR and in the range \pteeMax{8} at \five are presented in Fig.~\ref{fig:fits_pps}.

The data in the IMR are fitted in \mee and \ptee with PYTHIA 6 and POWHEG templates of open-charm (red) and open-beauty (magenta) production, keeping the light-flavor and \jpsi contributions fixed. In this mass range, most of the \ee~pairs originate from open heavy-flavor hadron decays. 
The $\chi^{2}$/ndf between the data and the cocktail sum is 110.9/123 for the POWHEG cocktail and 113.4/123 for the PYTHIA 6 cocktail. Both calculations are able to reproduce the measured spectra well over the full kinematic range probed, however the full cocktail obtained with POWHEG leads to a slightly better description of the data at low \mee around \mee = 0.5~\GeVcc.  The resulting cross sections are listed in Table~\ref{tab:crossSecs}. The systematic uncertainties originating from the data were determined by repeating the fit after moving the data points coherently up- and downward by their systematic uncertainties. 
Additional uncertainties on the effective beauty- and charm-to-electron branching ratios, arising from the semi-leptonic decay branching ratios of open heavy-flavor hadrons and the fragmentation functions of charm (beauty) quarks, amounting to 22\% and 6\% for the charm and beauty cross sections, respectively, are also listed in the table.
All uncertainties are fully correlated between the two generators, which differ only in the implementation of the heavy-quark production mechanisms. In both calculations, the hadronization of the c- and b-quarks, and the decays of the open heavy-flavor hadrons, are performed using PYTHIA 6. For the following results, only calculations where the heavy-flavor contribution is evaluated with POWHEG are presented, since the cocktail using POWHEG and fitted to the data in the IMR can slightly better describe the measured dielectron cross sections over the full \mee and \ptee range in \pp collisions at \five. 

\begin{table}[htb!]
    \centering
    \caption{Heavy-flavor cross sections extracted via double differential fits in \mee and \ptee to the measured dielectron spectra in \pp~collisions at \five using PYTHIA 6 and POWHEG. The statistical (stat.) and systematic (syst.) uncertainties on the data are quoted together with the 22\% (6\%) uncertainty on the branching ratio (BR) of the semi-leptonic decays of the open heavy-flavor hadrons and the fragmentation functions of charm (beauty) quarks.  The charm cross sections and corresponding uncertainties were updated in the recent submission. For details see Appendix~\ref{app:update}.}
    \renewcommand{\arraystretch}{1.3}
    \begin{tabular}{@{\extracolsep{0.3mm}} c c c }
    \hline
    \hline
         & PYTHIA & POWHEG \\
        \hline
        d$\sigma_{\rm c\overline{c}}$/d$y|_{y=0}$ & $\rm 900 \pm 105(stat) \pm 45(syst)$ $^{+208}_{-152}$ (BR),\ $\mu$b & $\rm 1299 \pm 137(stat) \pm 65(syst)$ $^{+300}_{-220}$ (BR)\,$\mu$b\\

        d$\sigma_{\rm b\overline{b}}$/d$y|_{y=0}$ & 34\,$\pm$\,4\,(stat.)\,$\pm$\,2\,(syst.)\,$\pm$\,2\,(BR)\,$\mu$b & 28\,$\pm$\,5\,(stat.)\,$\pm$\,1\,(syst.)\,$\pm$\,2\,(BR)\,$\mu$b\\
        \hline
        \hline
    \end{tabular}
    \label{tab:crossSecs}
\end{table}

\begin{figure}[htb]
    \begin{center}
    \includegraphics
    [width = 0.49\textwidth]{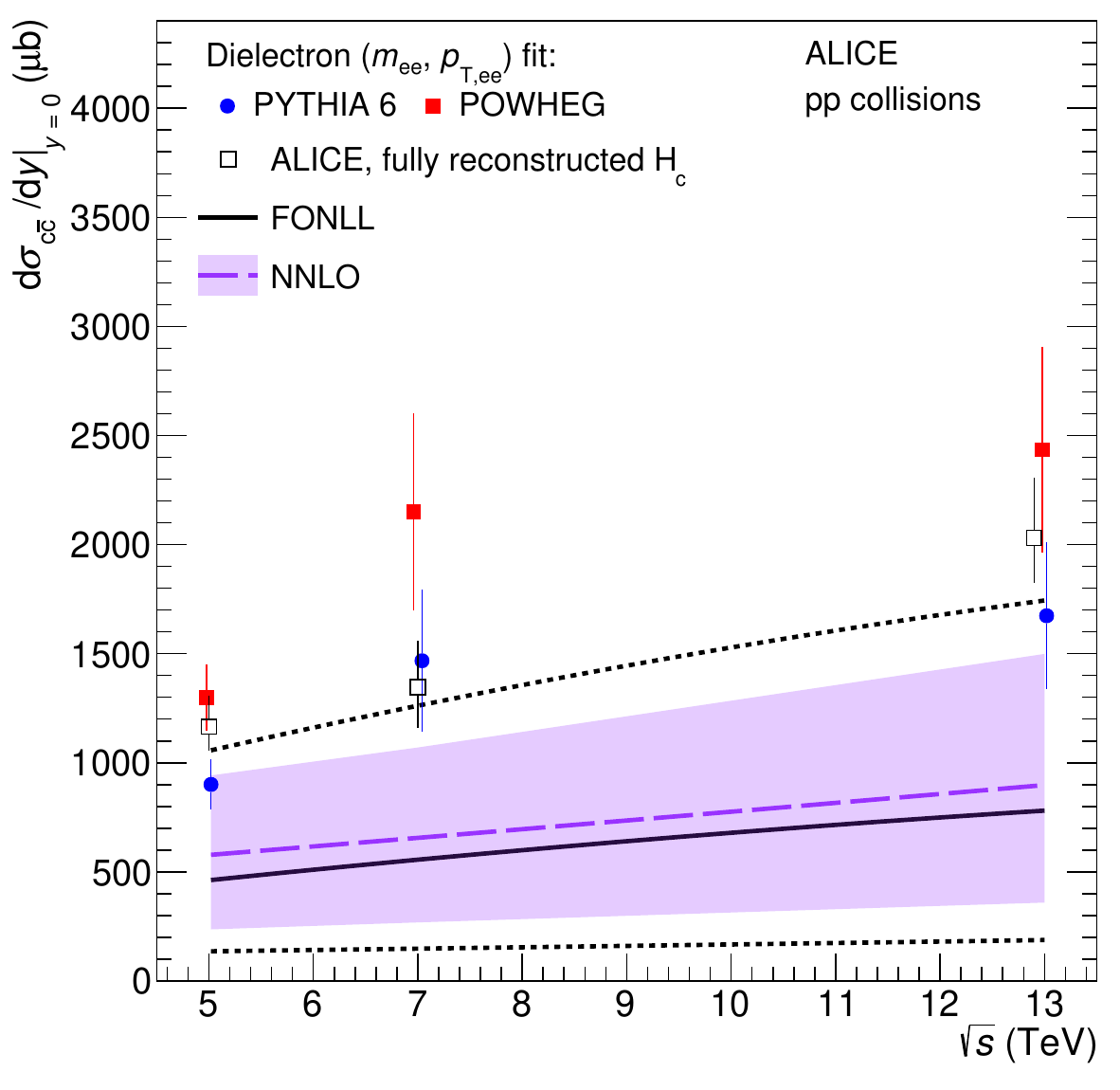}
    \includegraphics
    [width = 0.49\textwidth]{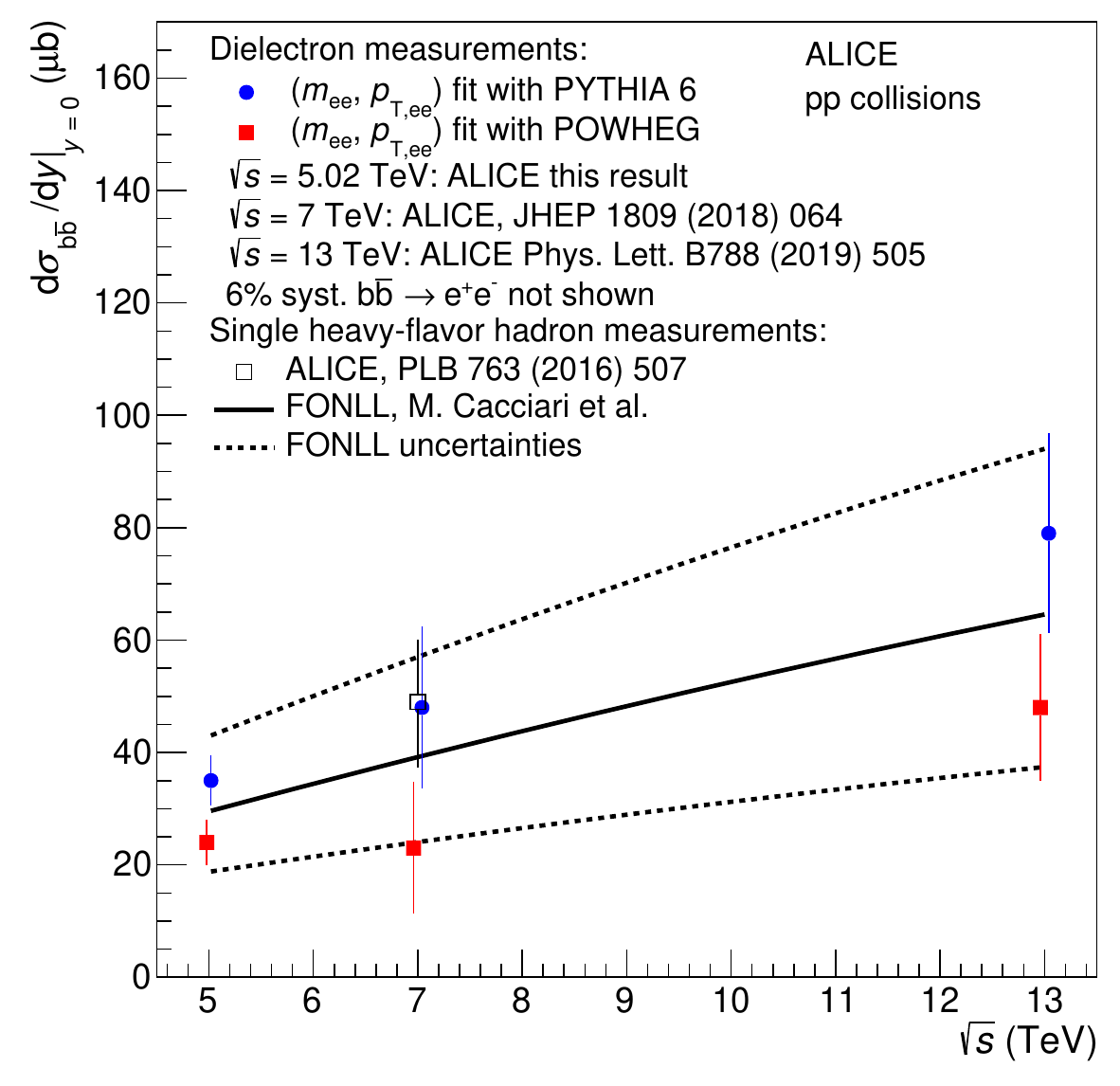}
    \end{center}
    \caption{(Color online) Cross sections at midrapidity for c$\overline{\textrm{c}}$ (left) and b$\overline{\textrm{b}}$ (right) as a function of $\sqrt{s}$ in \pp~collisions. The colored markers represent the measured midrapidity cross sections at \s = 5.02, 7, and 13 TeV which are derived using either PYTHIA 6 (blue circles) or POWHEG (red squares) simulations. The systematic and statistical uncertainty of the data points are summed in quadrature and represented by vertical bars. 
    The measurements are compared with FONLL calculations (black solid line), with model uncertainties (dashed lines),  and to single heavy-flavor hadron measurements (open markers).
    The charm cross sections and corresponding uncertainties were updated in the recent submission. For details see Appendix~\ref{app:update}. In addition to updating the \ccbar cross sections, recent measurements based on fully-reconstructed charm hadrons~\cite{ALICE:2021dhb, ALICE:2023sgl} were included in the figure as well as calculations at next-to-next-to-leading-order (NNLO) calculations (magenta)~\cite{dEnterria:2016ids, dEnterria:2016yhy}.
    }
    \label{fig:sqrt_comp}
\end{figure}

The compilation of the measured \ccbar and \bbbar cross sections as a function of $\sqrt{s}$ is shown in Fig.~\ref{fig:sqrt_comp} left and right, respectivly. In addition, complementary measurement of the cross sections at midrapidity in the fully reconstructed hadron channels in pp collisions at  are shown~\cite{ALICE:2021dhb,ALICE:2023sgl,ALICE:2012acz}. Within uncertainties the dielectron measurements are consistent with the results using single heavy-flavour hadrons. The measurements are compared to predictions calculated in the fixed-order next-to-leading-log (FONLL) approach~\cite{Cacciari:1998it, Cacciari:2001td, Cacciari:2012ny, Cacciari:2015fta}. Shifting up the data points coherently, the $\sqrt{s}$ dependence of the cross section is unchanged with respect to the original publication and described by the FONLL calculations. The model uncertainties are dominated by scale uncertainties, but also include PDF and mass uncertainties.  The slope of the \mbox{center-of-mass} energy dependence of the cross sections is described by the FONLL calculations. While consistent with FONLL predictions within uncertainties, the measurements of the \ccbar cross sections tend to be on the upper edge of the theoretical uncertainty band. Calculations at next-to-next-to-leading-order (NNLO) predict a consistently higher \ccbar cross section as a function of $\sqrt{s}$~\cite{dEnterria:2016ids, dEnterria:2016yhy}. The smaller uncertainties however increase the tension with the measurements.
\footnote{The discussion of the results was updated in the recent submission to adapt changes to the original. Further discussions of the updated \ccbar cross sections and figure \ref{fig:sqrt_comp} (left) can be found in Appendix~\ref{app:update}.}

\subsection{Dielectron production in \pp~and \pPb~collisions}
\label{resultsPPB}

\begin{figure}[htb]
\begin{center}
    \includegraphics
    [scale= 0.39]{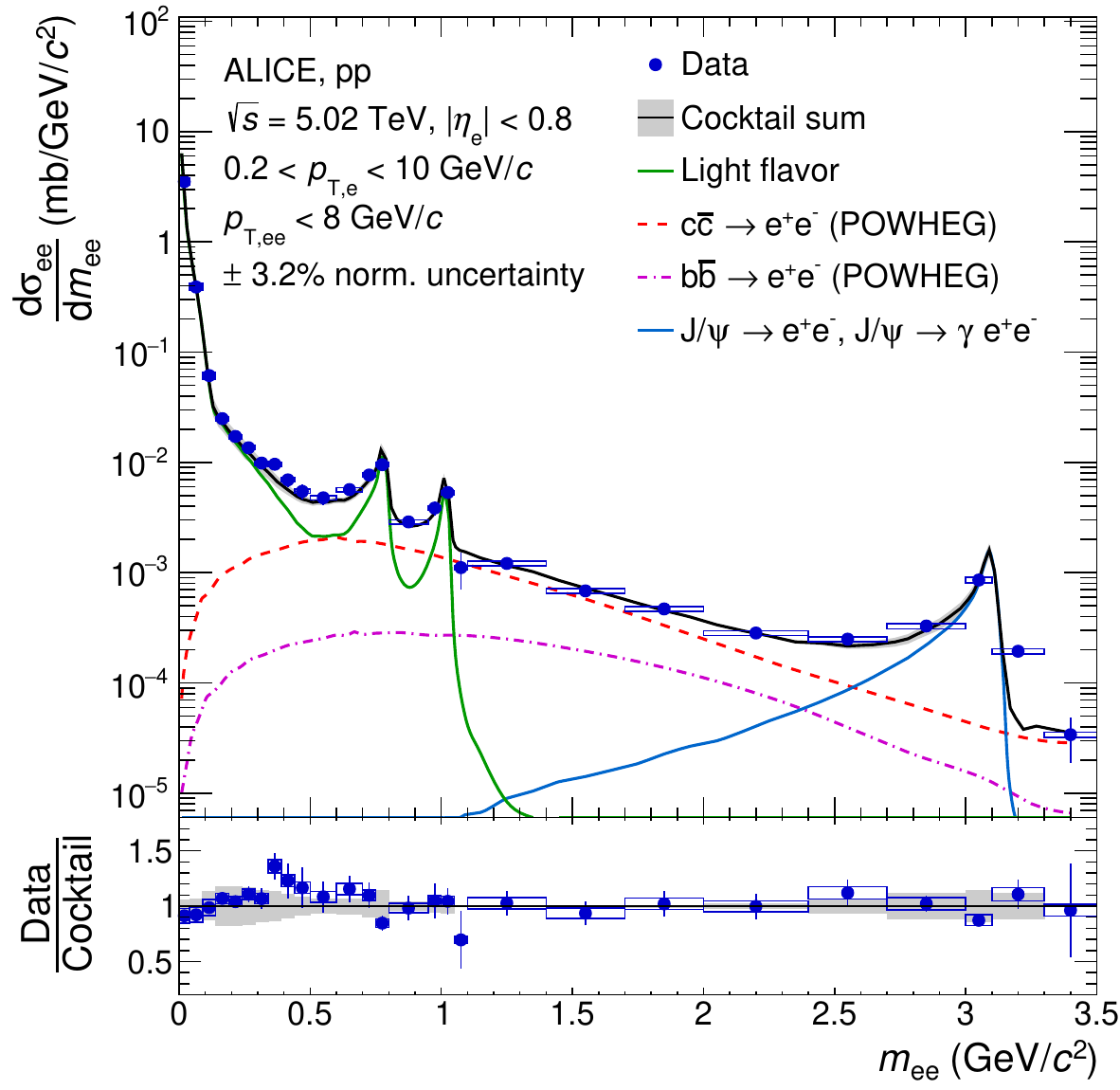}
    \includegraphics
    [scale= 0.39]{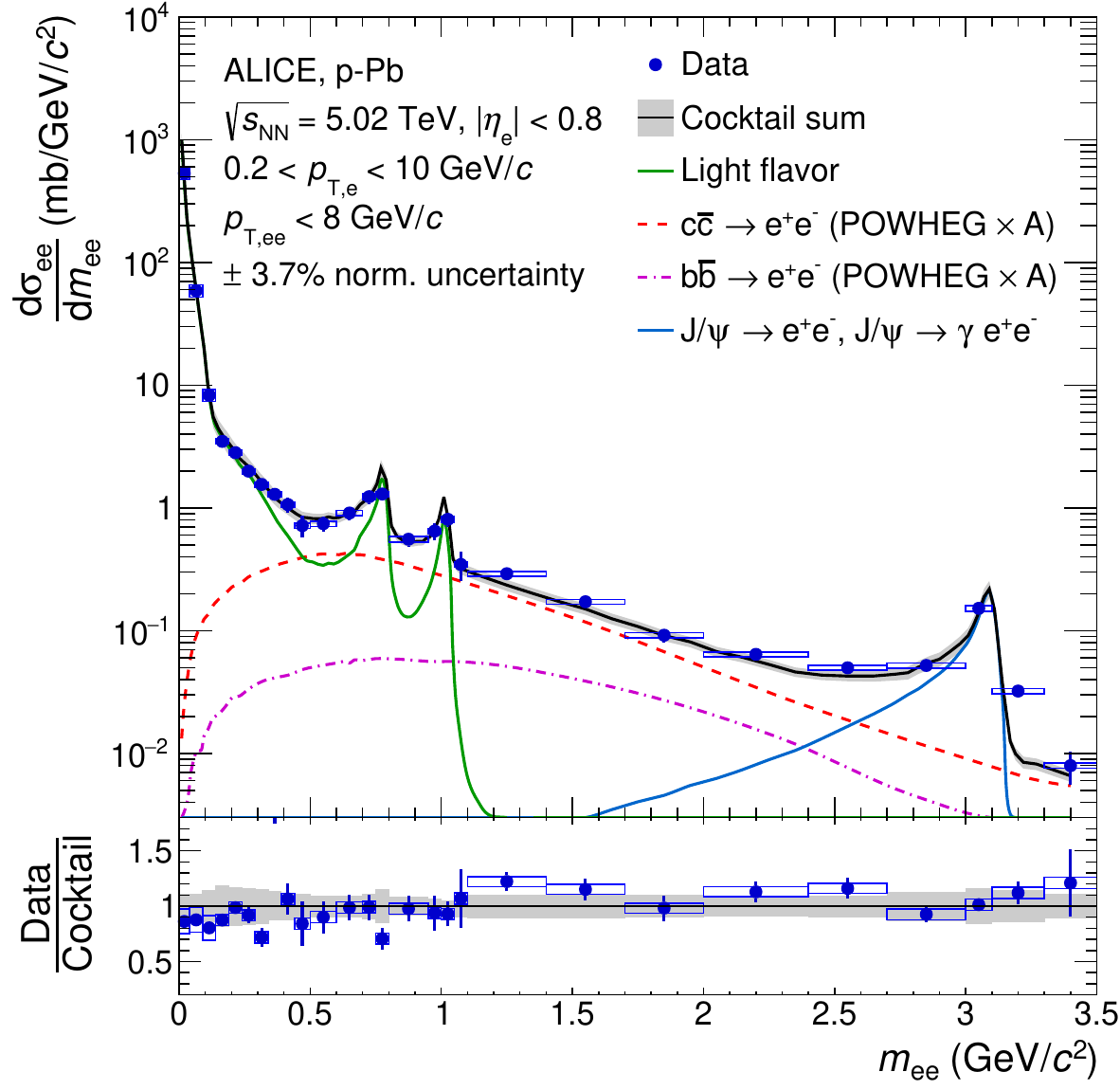}
    \end{center}
     \caption{(Color online) Differential \ee~cross section as a function of \mee measured in \pp~(left) and \pPb~(right) collisions at \fivenn. The data are compared to the hadronic cocktail, where the heavy-flavor contributions are fitted to the \pp~spectrum in the intermediate-mass region, and for \pPb~collisions scaled with the atomic mass number of the Pb nucleus $A$ = 208. The gray band represents the total uncertainty on the hadronic cocktail.}
    \label{fig:mee}
\end{figure}

The \mee-differential production cross sections of \ee~pairs measured in \pp~and \pPb~collisions at \fivenn~are compared to the expected dielectrons from known hadron decays in Fig.~\ref{fig:mee}. The light-flavor contributions, summarized as "Light flavor" for readability, are based on measurements in \pp and \pPb~collisions as explained in detail in Sec.~\ref{cocktail}.
The correlated pairs from heavy-flavor hadron decays are calculated with POWHEG. Their contributions are normalized to the d$\sigma_{\rm c\overline{c}}$/d$y|_{y=0}$ and the d$\sigma_{\rm b\overline{b}}$/d$y|_{y=0}$ in \pp~collisions obtained from the fit to the \pp~data, as discussed in the previous section. For \pPb~collisions, the heavy-flavor contributions are further scaled with the atomic mass number of the Pb nucleus. This assumes that the production of heavy-flavor quarks in \pPb collisions scales with the number of binary nucleon--nucleon collisions. The total systematic uncertainty of the cocktails is indicated by the gray band. The \pp~cocktail uncertainty in the IMR is zero by construction since the heavy-flavor contribution is directly fitted to the measured spectrum in \pp~collisions. 
The systematic uncertainties of the heavy-flavor contribution in the \pPb~cocktail originate from the statistical and systematic uncertainties of the extracted production cross sections in the \pp~analysis listed in Table~\ref{tab:crossSecs}.
Since the cross section is based on the measurement of final state \ee~pairs, the uncertainties related to branching ratios of the semi-leptonic decays of open heavy-flavor hadrons and the fragmentation functions of charm and beauty quarks can be omitted, under the assumption that these do not change from \pp~to \pPb~collisions. This is confirmed by the latest measurements of open heavy-flavor hadrons in \pp~and \pPb~collisions at \fivenn by ALICE~\cite{Acharya:2017kfy}. The bottom panels in Fig.~\ref{fig:mee} show the ratios of the data to the cocktail. 
The data are described by the hadronic cocktails over the whole mass range (\meeMax{3.5}) in both \pp~and \pPb~collisions, within the systematic and statistical uncertainties. As seen in previous measurements in \pp~collisions~\cite{pp7tev,pp13tev}, the heavy-flavor contribution dominates the spectrum for \meeMin{0.8}. In \pPb~collisions, the heavy-flavor contribution to the hadronic cocktail does not include any modification beyond scaling with binary nucleon--nucleon collisions with respect to the \pp~cocktail. No significant deviation of the data from the vacuum expectation of the heavy-flavor contributions can be observed in the mass spectrum. This suggests that the CNM effects are small compared to the current uncertainties of the measurements, as observed by other open heavy-flavor measurements at the LHC at midrapidity~\cite{Dmeson2019}, or compensated by an additional source of dielectrons in \pPb collisions compared with \pp collisions, possibly related to the formation of a hot medium in such collisions.

\begin{figure}[htb]
    \begin{center}
    \includegraphics
    [width = 0.49\textwidth]{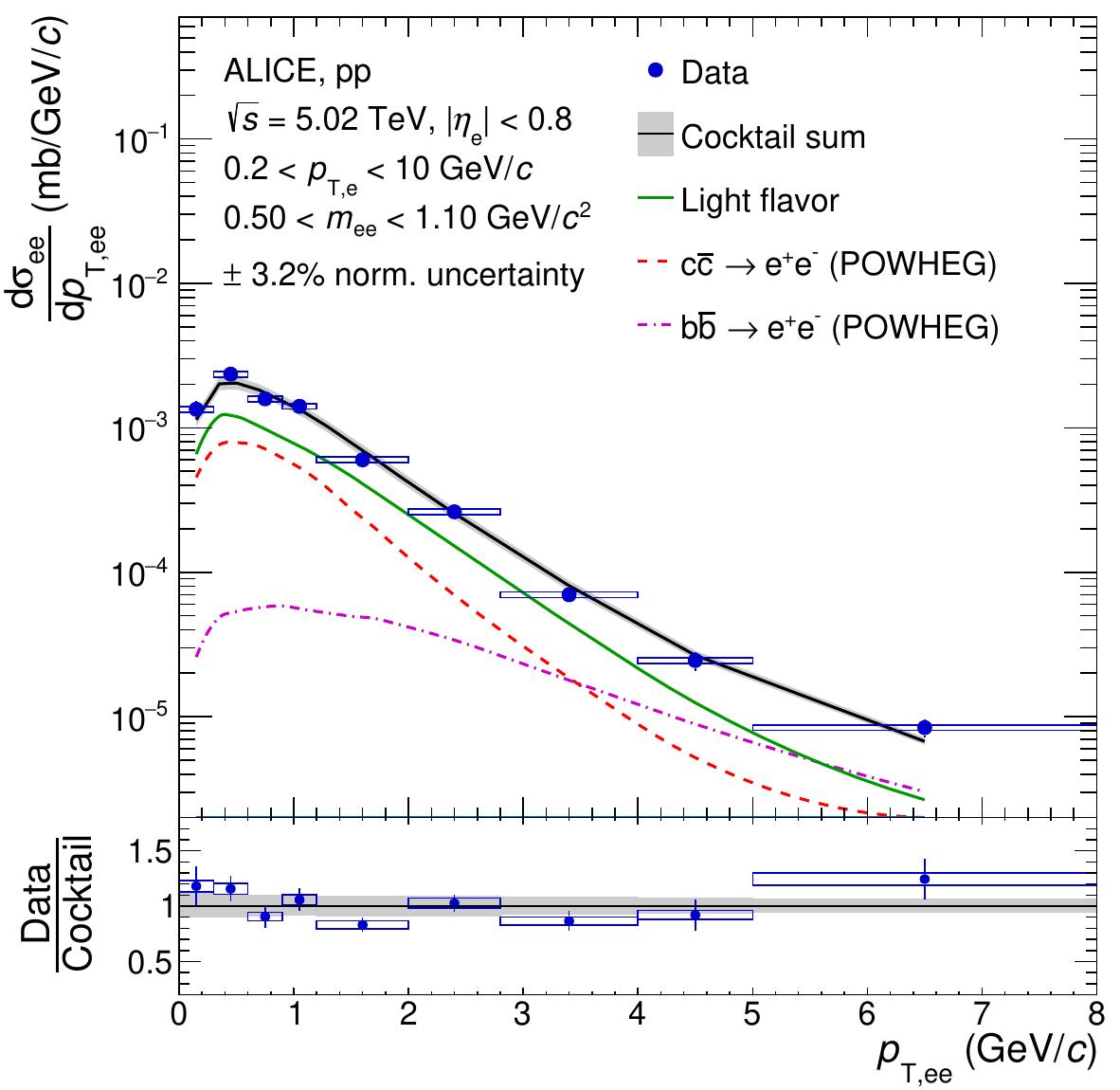}
    \includegraphics
    [width = 0.49\textwidth]{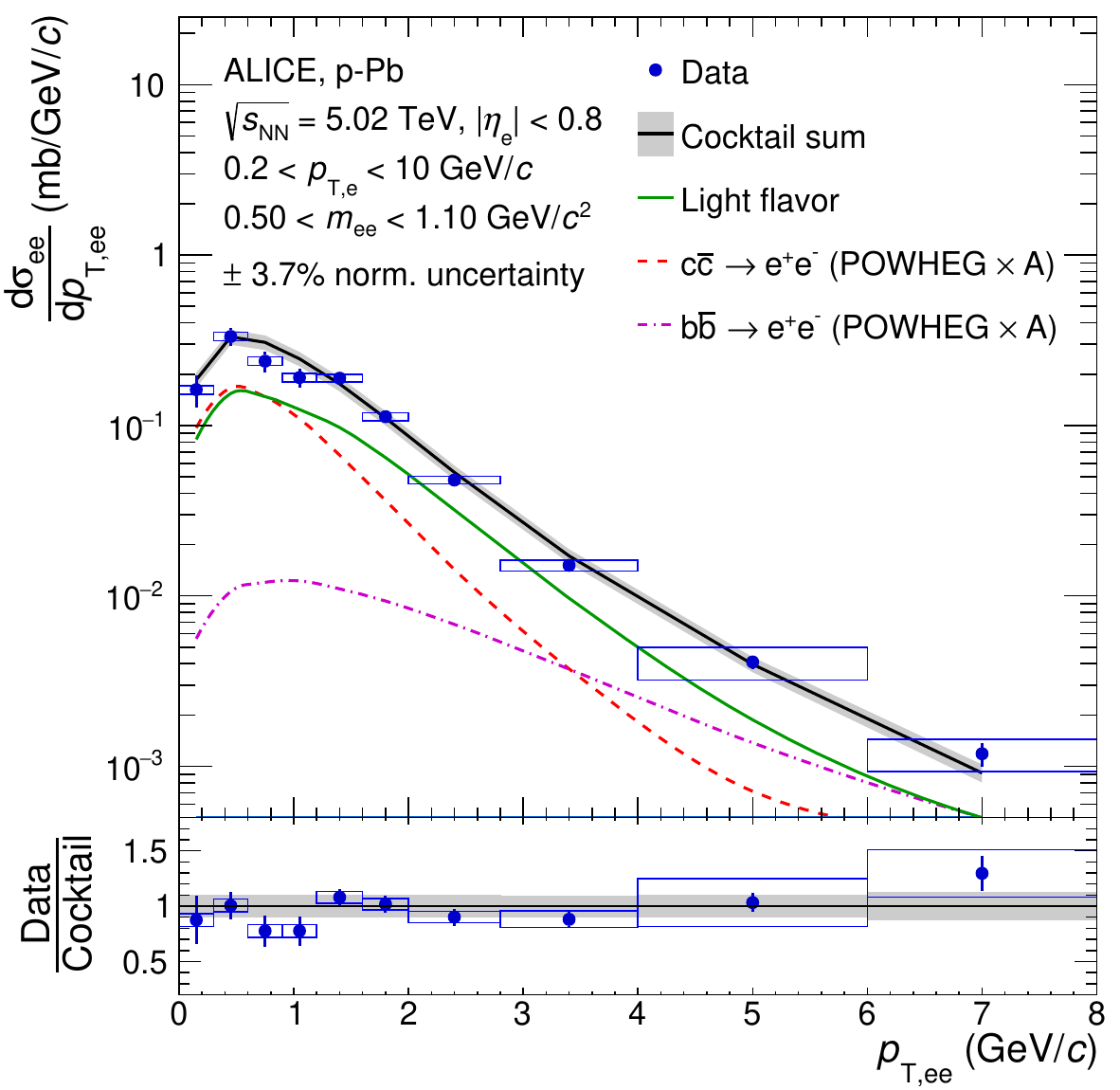}
    \end{center}
    \caption{(Color online) Differential \ee~cross section as a function of \ptee in the low-mass region measured in pp (left) and \pPb~(right) collisions at \fivenn. The data are compared to the hadronic cocktail, where the heavy-flavor contributions are fitted to the \pp~spectrum in the intermediate-mass region, and for \pPb~collisions scaled with the atomic mass number of the Pb nucleus $A$ = 208. The gray band represents the total uncertainty on the hadronic cocktail.}
    \label{fig:ptee_resonance}
\end{figure}

\begin{figure}[htb]
    \begin{center}
    \includegraphics
    [width = 0.49\textwidth]{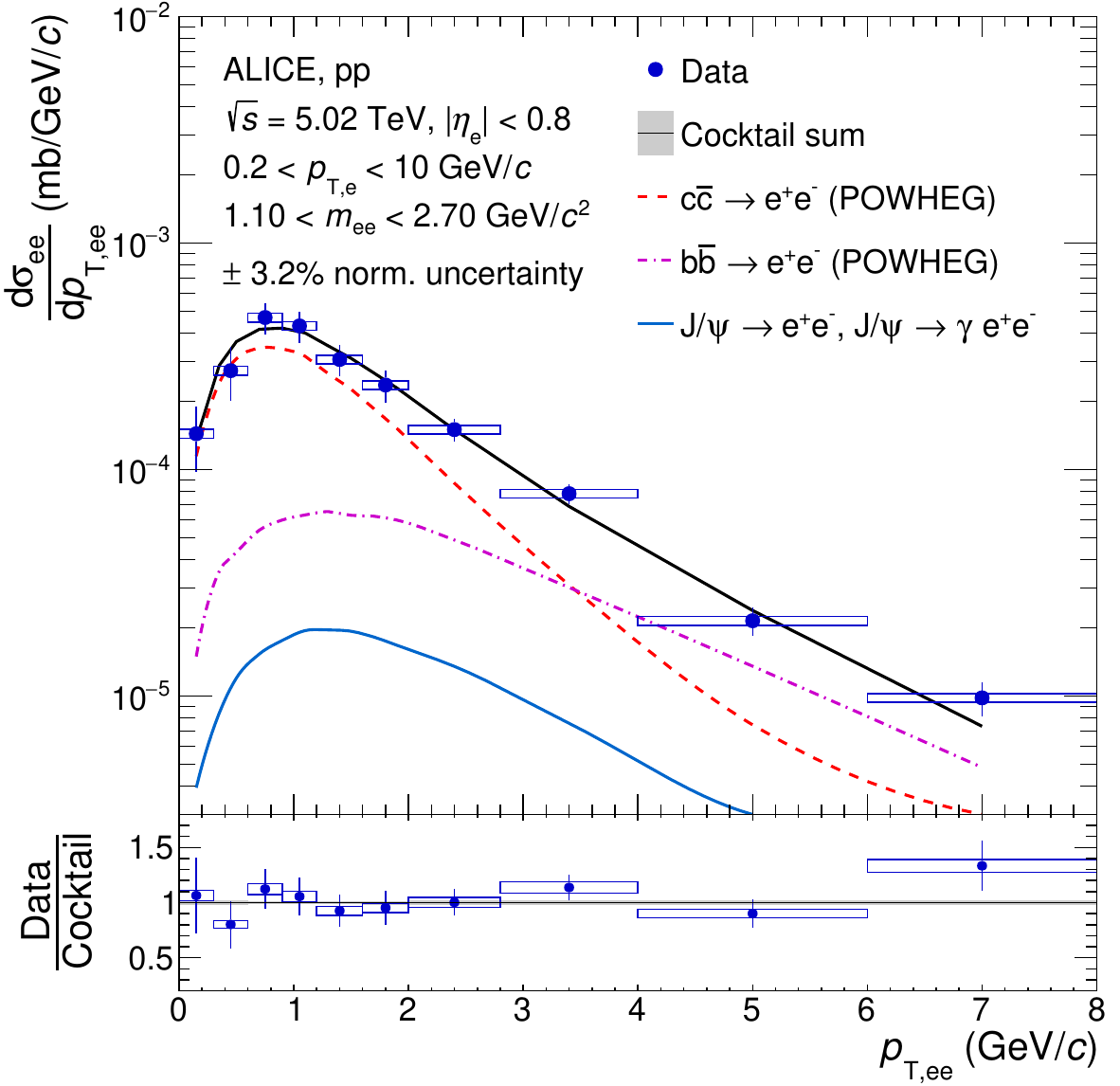}
    \includegraphics
    [width = 0.49\textwidth]{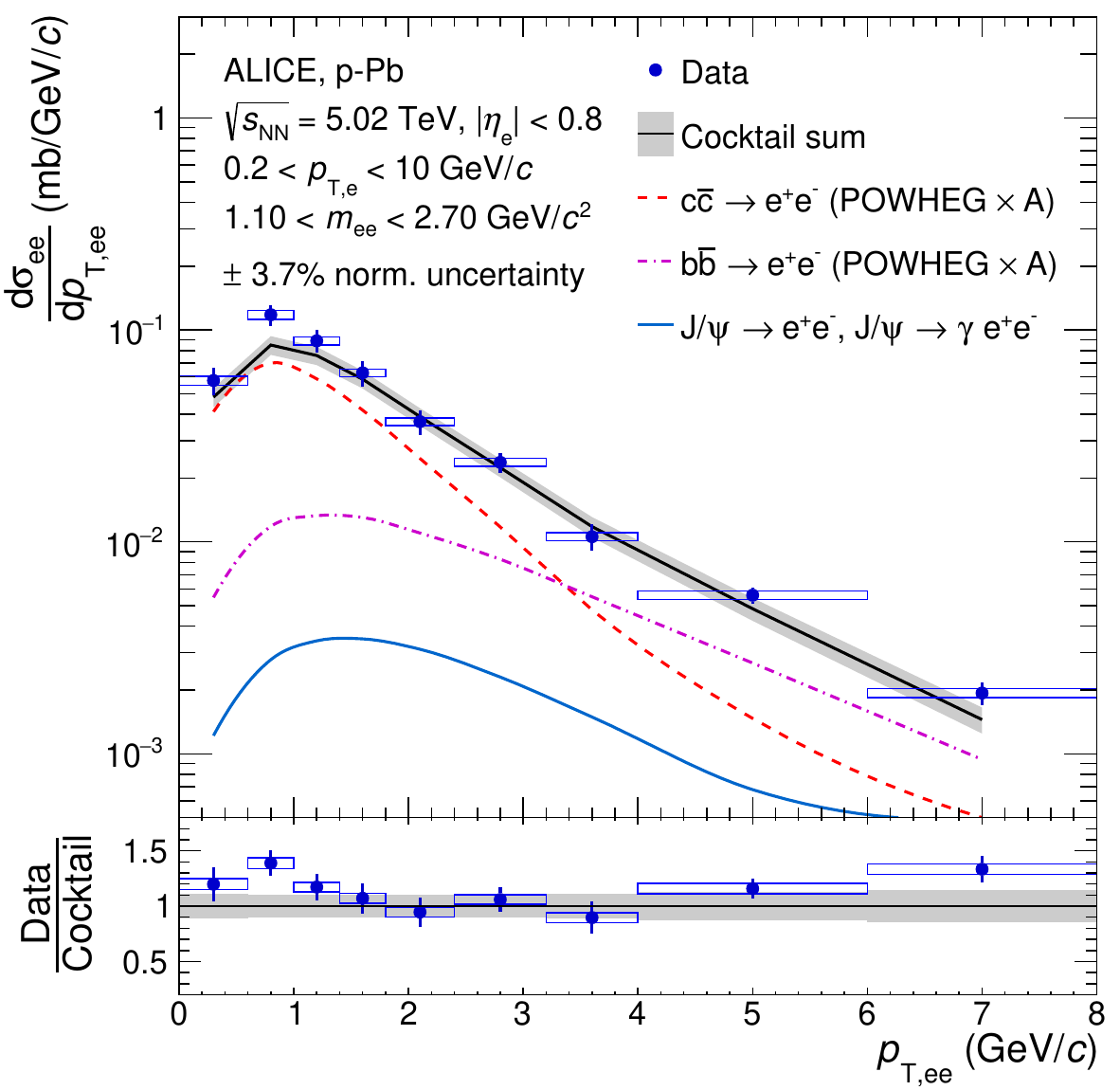}
    \end{center}
    \caption{(Color online) Differential \ee~cross section as a function of \ptee in the intermediate-mass region measured in pp (left) and \pPb~(right) collisions at \fivenn. The data are compared to the hadronic cocktail, where the heavy-flavor contributions are fitted to the \pp~spectrum in the intermediate-mass region, and for \pPb~collisions scaled with the atomic mass number of the Pb nucleus $A$ = 208. The gray band represents the total uncertainty on the hadronic cocktail.}
    \label{fig:ptee_IMR}
\end{figure}

The \ptee spectra for \pp~and \pPb~collisions in the LMR and IMR are compared to the hadronic cocktail in Figs.~\ref{fig:ptee_resonance} and~\ref{fig:ptee_IMR}, respectively.
In the LMR, the hadronic cocktails in \pp and \pPb collisions are both composed of \ee~pairs from light-flavor, open-charm, and open-beauty hadron decays. Most of the pairs in this mass interval are produced from the decays of light-flavor hadrons, whose production at low \pt does not scale with $A$ in \pPb~collisions. Therefore, the relative expected contribution of dielectrons from light-flavor hadron decays is smaller in \pPb~collisions compared with \pp~collisions at the same $\sqrt{s_{\rm NN}}$. In \pPb~collisions, the open-charm hadron decays are expected to contribute significantly to the \ee~cross section for \pteeMax{1}. The open-beauty contribution only plays a significant role for \pteeMin{4} in both collision systems. 
In the IMR, correlated \ee~pairs from open-charm hadron decays are the dominant dielectron source for \pteeMax{2.5} in \pp as well as in \pPb collisions, whereas most of the \ee~pairs originate from open-beauty hadron decays for \pteeMin{3.5}. The contribution from \jpsi decays is small over the whole \ptee range.
The dielectron production in \pp~and \pPb~collisions at \fivenn is well described by the hadronic cocktail, utilizing heavy-flavor cross sections fitted to the \pp~data and assuming a scaling of the heavy-flavor cross sections with the $A$ of the Pb nucleus. In particular, no significant modification of the heavy-flavor production in the measured kinematic regions is justified by the analysis of the \pPb collisions data.

\subsection{Nuclear modification factor}

\begin{figure}[htb]
    \begin{center}
    \includegraphics[width=0.49\textwidth]{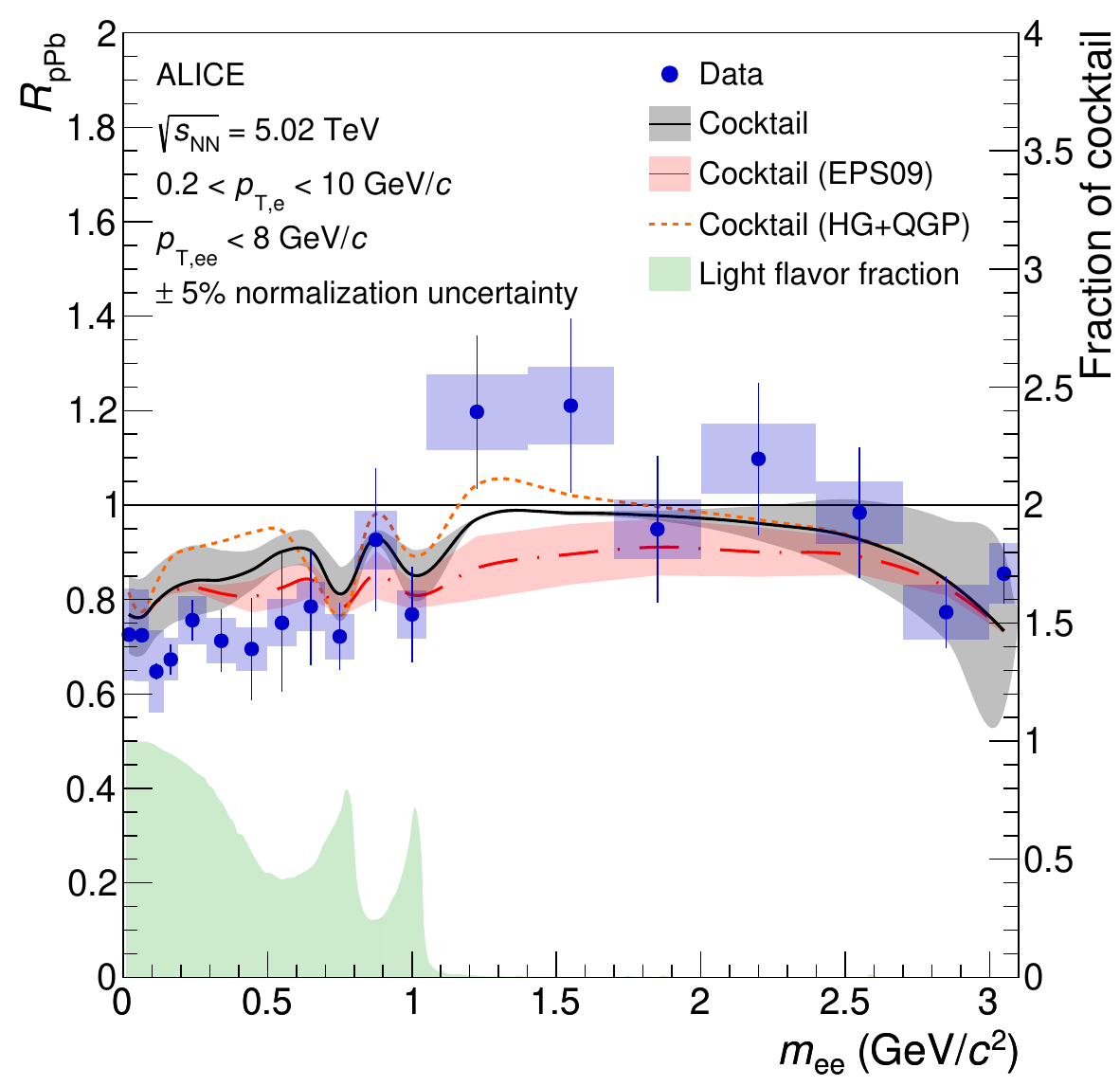}
    \end{center}
    \caption{(Color online) Measured dielectron nuclear modification factor as a function of \mee at \fivenn. The data are shown in blue, with their statistical and systematic uncertainties depicted as vertical bars and boxes. The baseline expectation, calculated from the \pp~and \pPb~cocktails outlined in Sec.~\ref{cocktail}, is shown as a black line with a gray band indicating its uncertainties. Two additional cocktails, one incorporating a modified charm production due to CNM effects and another one including thermal radiation from the hadronic and partonic phases, are shown as red and orange dashed lines, respectively.}
    \label{fig:rppb-mass}
\end{figure}

The nuclear modification factor, \RpPb, is calculated as
\begin{equation}
    R_{\rm pPb}(\textit{m}_{\rm{ee}}) = \frac{1}{A}
    \frac{\rm{d} \sigma^{\kern+.1empPb}_{ee}/\rm{d}\textit{m}_{\rm{ee}}}
    {\rm{d} \sigma^{\kern+.1empp}_{\rm ee}/\rm{d}\textit{m}_{\rm{ee}}},
    \label{eq:nucMod}
\end{equation}
with $\sigma^{\kern+.1em\rm{pPb}}_{\rm ee}$ and $\sigma^{\kern+.1em\rm{pp}}_{\rm ee}$ representing the cross sections of dielectron production in \pPb and \pp~collisions, respectively, and $A$ denoting the mass number of the Pb nucleus (208). The \RpPb allows for a direct comparison of the measurements in the \pp~and \pPb~collision systems. The systematic uncertainties of the \pPb~and \pp~measurements are treated as independent and, thus, added in quadrature. The dielectron \RpPb as a function of \mee for \pteeMax{8} is shown in Fig.~\ref{fig:rppb-mass}.
The data are compared to the \RpPb of the hadronic cocktails as described in Sec.~\ref{cocktail} (solid black line). In the cocktail \RpPb, the uncertainties from the open heavy-flavor contributions as well as those from the scaling factor applied to the $\pi^{\pm}$ parametrizations, the $\rho/\pi^{\pm}$, $\omega/\pi^{\pm}$, and $\eta/\pi^{0}$ \pt ratios are fully correlated, and therefore cancel out. The uncertainties on the parametrized $\pi$, $\phi$, and \jpsi spectra are propagated to the \RpPb. Since they are based on independent measurements they are added quadratically. The measured \RpPb is below the expectation of binary collision scaling for \meeMax{1.1}, where the fraction of dielectrons from light-flavor hadron decays to the total expected \ee~cross section in \pPb~collisions, denoted by the green area,  is not negligible. The \RpPb is consistent with unity in the IMR within uncertainties, displaying a step between the two mass regions. The behavior is reproduced, within uncertainties, by the hadronic cocktail assuming no further modification of the open heavy-flavor cross sections beyond binary collision scaling. This suggests a different scaling behavior of the light-flavor production from binary collision scaling, as already indicated in previous measurements~\cite{Acharya:2018qsh}. 

An additional cocktail calculation incorporating a modification of the open-charm contribution via CNM effects is shown by a dashed red line in Fig.~\ref{fig:rppb-mass}. 
The CNM effects on the production of dielectrons from open-charm hadron decays are incorporated by using the EPS09 nPDF~\cite{EPS09} in the POWHEG calculations. 
In the mass region below 1~\GeVcc, where the admixture of charm is significant, the modification of the charm contribution improves the description of the measured $R_{\rm pPb}$. In the IMR, the data are just beyond the upper edge of the systematic uncertainties of the calculations including CNM for the charm production. On one hand, it suggests negligible CNM effects compared to the current precision of the measurement in this mass range, where the \pt of D mesons, from which the dielectrons originate, is larger than 2~\GeVc according to calculations performed with PYTHIA 6. This is in agreement with previous results on the D meson \RpPb at \fivenn by ALICE, which show no significant modification of the \pt spectra above 2~\GeVc~\cite{Dmeson2016,Dmeson2019} compared to \pp~collisions. The dielectron cross section from charm at lower \mee however is sensitive to the production of low \pt D mesons (\pt $<$ 2~\GeVc). On the other hand, a possible additional source of electron pairs in \pPb collisions compared to \pp collisions could compensate CNM effects on the heavy-flavor production. 

The measured \RpPb is further compared to calculations including thermal radiation from the hadronic and partonic phases, based on a model which describes the dilepton enhancement measured in heavy-ion collisions at the SPS and RHIC~\cite{Rapp:2000pe,RAPP200013,PhysRevLett.96.162302,Rapp:2013nxa}. The contribution of thermal dielectrons is obtained from an expanding thermal fireball model for \pPb collisions at \fivenn, corresponding to a mean charged-particle multiplicity at midrapidity of $\langle {\rm d}N_{\rm ch}/{\rm d}y \rangle$ = 20, corrected for weak decay feeddown. The equation of state was extracted from lattice QCD computations with a crossover transition around the critical temperature $T_{\rm c}$ = 170\,MeV. A broadening of the $\rho$ electromagnetic spectral function is expected as an effect of interactions in the hot hadronic phase. The thermal emission rate of dielectrons from the hadronic phase is calculated based on the hadronic many-body theory. 
The effects of the detector resolution are not included in the calculations and no modification of the heavy-flavor contribution is considered. A hadronic cocktail including these calculations is shown as the orange dotted line (HG+QGP). In the range \meeRange{0.2}{0.6}, the model tends to slightly overestimate the measured \RpPb, whereas in the IMR it agrees with the data within their uncertainties. An additional thermal source of dielectrons in \pPb collisions compared to \pp collisions can not be excluded by the data.

\begin{figure}[htb]
    \begin{center}
    \includegraphics[width = 0.49\textwidth]{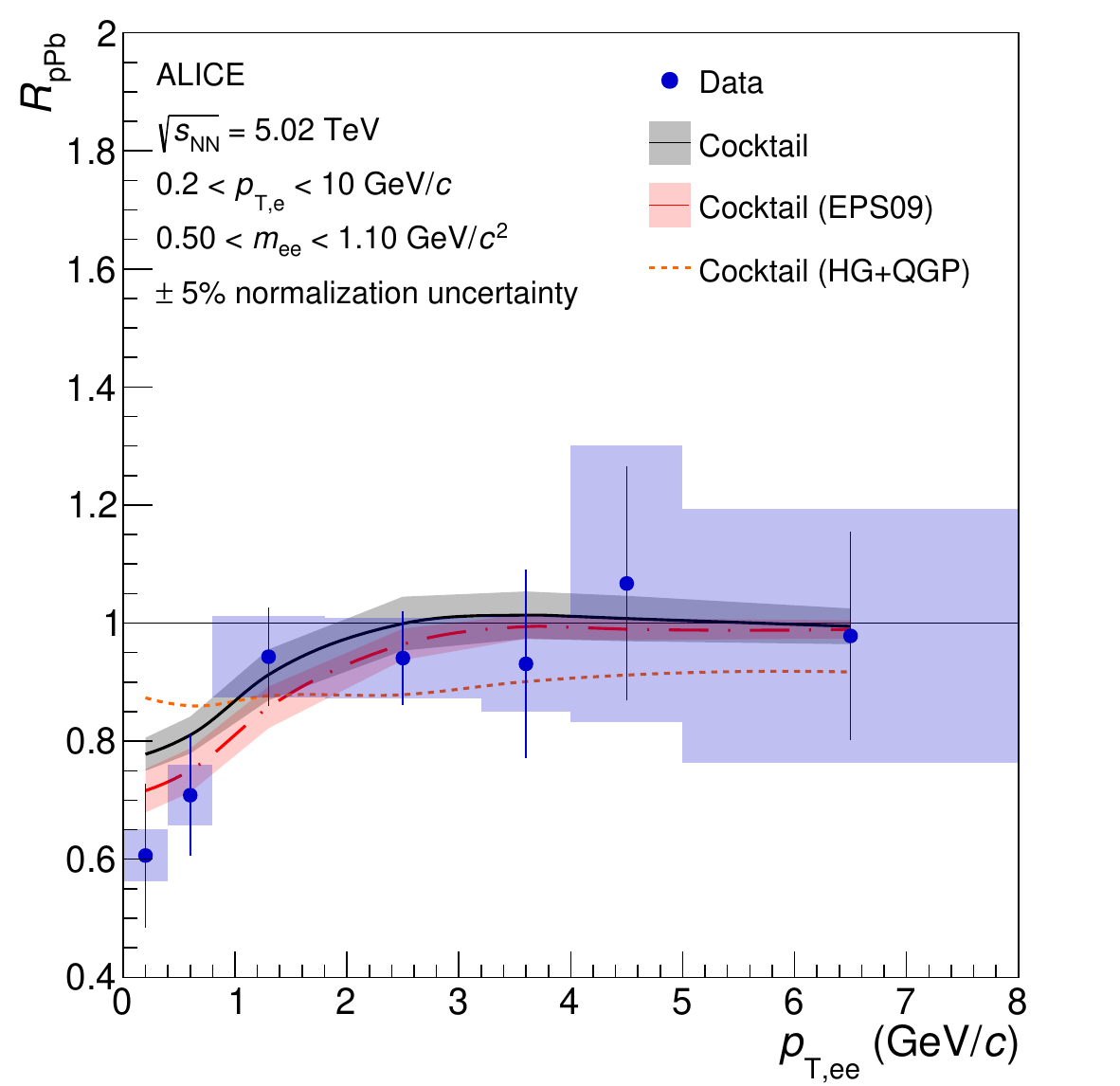}
    \includegraphics[width = 0.49\textwidth]{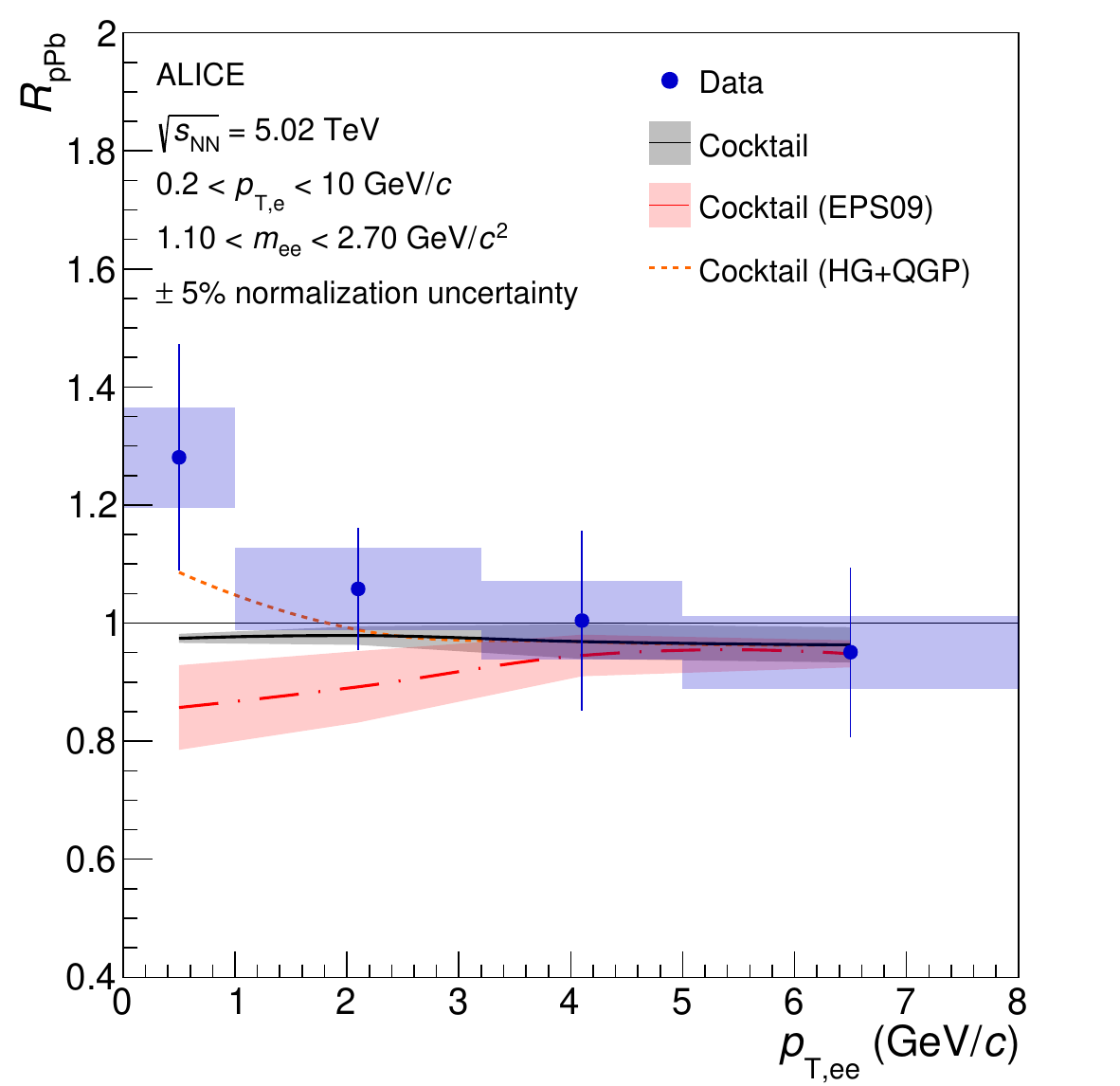}
    \end{center}
    \caption{(Color online) Measured dielectron nuclear modification factor as a function of \ptee in the low-mass region (left) and intermediate-mass region (right) at \fivenn. The data are shown in blue, with their statistical and systematic uncertainties depicted as vertical bars and boxes. The baseline expectation, calculated from the \pp~and \pPb~cocktails outlined in Sec.~\ref{cocktail}, is shown as a black line with a gray band indicating its uncertainties. Two additional cocktails, one incorporating a modified charm production due to CNM effects and another one including thermal radiation from the hadronic and partonic phases, are shown as red and orange dashed lines, respectively.}
    \label{fig:rppb-pt}
\end{figure}

To further investigate the modifications of the open-charm contribution to the \ee~spectrum, the dielectron \RpPb as a function of \ptee~is shown in the LMR and IMR in Fig.~\ref{fig:rppb-pt}.

In the LMR, the fraction of \ee~pairs from light-flavor hadron decays ranges from about 40\% to 60\% depending on \ptee. For \ptee larger than about 1~\GeVc the data are compatible with binary collision scaling, indicating that the production of light-flavor hadrons is driven by the initial hard scatterings of the incoming partons and is not affected by CNM effects.
This no longer holds true for \pteeMax{1}, pointing to a change in the production mechanism of the light-flavor hadrons. These features can be reproduced by the hadronic cocktail.
Inclusion of CNM effects for the charm contribution in the hadronic cocktail only have a small effect. The uncertainties on the data as well as the CNM calculations themselves are too large to draw any conclusion.
The addition of the thermal contributions in the LMR is disfavored by the data at low-\ptee (\pteeMax{1}), whereas at higher \ptee the uncertainties on the data do not allow for any discrimination between the three models.

In the IMR, the contribution from light-flavor hadron decays is negligible. The \RpPb is consistent with unity, indicating that the heavy-flavor cross sections approximately scale with the number of binary collisions in this range. According to the calculations using EPS09 nPDFs, a suppression of the total \ee~cross section is expected for \pteeMax{3.5} due to CNM effects on dielectrons from open-charm hadron decays. Nevertheless, it is disfavored by these data. 
On the contrary, the cocktail calculation including thermal contributions would be preferred by the data. In particular, for \pteeMax{1} a thermal contribution significantly helps to improve the description of the data.

Finally, a potential interplay between CNM effects and the thermal contribution cannot be ruled out. Therefore, it is mandatory to separate the dielectrons from heavy-flavor hadron decays and those from thermal radiation. This could be achieved by an analysis as a function of the distance-of-closest approach of the \ee~pairs to the collision vertex~\cite{pp7tev}.

\section{Conclusions}
\label{chap:conclusion}

The dielectron production at midrapidity (\etarangee{0.8}) was measured with the ALICE detector as a function of invariant mass and pair transverse momentum in \pp~and \pPb~collisions at \fivenn.

In \pp~collisions, the dielectron continuum can be well described by the expected contributions from light-flavor hadron decays and calculations of \ee~pairs from heavy-flavor hadron decays fitted to the data. The cross sections of \ccbar and \bbbar production at midrapidity are extracted from the measurement by a double-differential fit to the \mee and \ptee spectrum in the intermediate-mass region. Templates from two different event generators, PYTHIA 6~\cite{pythia64} and POWHEG~\cite{powheg1,powheg2,powheg3,powheg4}, are used. Both calculations can describe the data well, yet they yield significantly different results for the cross sections of the single \ccbar and \bbbar contributions. The hadronization of c- and b-quarks as well as the decay of the heavy-flavor hadrons is done in both PYTHIA 6 and POWHEG simulations with the Perugia 2011 tune of PYTHIA~6.4~\cite{pythia64,perugia2011}. Therefore, the model dependence of the extracted cross sections directly reflects the sensitivity of the dielectron measurement to the different implementation of the heavy-quark production mechanisms in the Monte Carlo event generators. The measured ${\rm d}\sigma_{\rm c\overline{c}}/{\rm d}y|_{y=0}$ and ${\rm d}\sigma_{\rm b\overline{b}}/{\rm d}y|_{y=0}$ are compared to existing results from dielectron measurements, as well as measurements of identified charm hadrons and semi-leptonic decays of beauty hadrons, in \pp~collisions at different \s. The difference between the cross sections extracted in this analysis with the two event generators is comparable to those reported in previous observations at \s = 7 and 13~TeV. The slope of the \mbox{center-of-mass} energy dependence of the cross sections can be described by FONLL calculations. 

The dielectron \mee and \ptee spectra in \pPb~collisions at \fivenn, reported here for the first time, are compared to a hadronic cocktail composed of the expected dielectron cross sections from the known hadron decays. Whereas \ee~pairs from light-flavor and \jpsi hadron decays are estimated using independent measurements of hadrons, the contributions of dielectrons from open heavy-flavor hadron decays are determined from the dielectron measurement in \pp~collisions at the same \mbox{center-of-mass} energy using POWHEG as the event generator. The heavy-flavor cross sections are assumed to scale with the atomic mass number of the Pb nucleus in \pPb~collisions, with respect to the measured pp reference. Good agreement is observed between the measured and expected total \ee~cross section.

The dielectron \RpPb as a function of \mee highlights the different scaling behavior of the light- and heavy-flavor dielectron sources. While the measured \RpPb is below one for \meeMax{1}, it is consistent with unity within uncertainties in the IMR where most of the \ee~pairs originate from correlated open heavy-flavor hadron decays. On the one hand, calculations including a suppression of the charm production using the nPDF EPS09 do not describe the data as well as the hadronic cocktail using the atomic mass number scaling hypothesis in the intermediate-mass region. The central value of the computations including CNM effects is nevertheless closer to the measured \RpPb at masses around 0.5~\GeVcc. On the other hand, including a thermal contribution from a hot hadronic and partonic phase to the dielectron cocktail helps in the description of the data in the IMR. The thermal radiation calculations seem however to overestimate the production of dielectrons in the LMR. The hadronic cocktail calculations including CNM effects and thermal radiation show that both play a role at low \ptee with opposite trends, 
although the current uncertainties on the measured \ptee dependence of \RpPb are still too large to reject any of the calculations presented. Moreover, CNM effects on the charm production and thermal radiation from a hot medium possibly formed in \pPb collisions could cancel each other, if both are present, which makes it necessary to disentangle them in a more sophisticated approach.

A more detailed study of the dielectron production in \pPb~collisions requires the separation of \ee~pairs from prompt sources and those from the displaced open heavy-flavor hadron decays. The distance-of-closest approach of the \ee~pair to the collision vertex, pioneered at the LHC by ALICE in the dielectron analysis of the \pp~data at $\sqrt{s} =7$~\TeV~\cite{pp7tev}, could enable the search for the presence of a possible additional contribution from thermal radiation in \pPb~collisions, in particular in high-multiplicity events. 
In the near future, the dielectron analysis will greatly benefit from the upgrades of the ALICE TPC~\cite{CERN-LHCC-2013-020,CERN-LHCC-2015-002}, the ITS~\cite{Musa:1475244} and a completely new readout system~\cite{newReadoutALICE} and computing framework~\cite{newCompFrameworkALICE}. The data acquisition rate will increase by a factor of 100, while the pointing resolution of primary tracks will improve by a factor of 3 to 6, depending on their orientation with respect to the magnetic field. This will open up the possibility to study the dielectron production with unprecedented precision and detail.

\newenvironment{acknowledgement}{\relax}{\relax}
\begin{acknowledgement}
\section*{Acknowledgements}

The ALICE Collaboration would like to thank all its engineers and technicians for their invaluable contributions to the construction of the experiment and the CERN accelerator teams for the outstanding performance of the LHC complex.
The ALICE Collaboration gratefully acknowledges the resources and support provided by all Grid centres and the Worldwide LHC Computing Grid (WLCG) collaboration.
The ALICE Collaboration acknowledges the following funding agencies for their support in building and running the ALICE detector:
A. I. Alikhanyan National Science Laboratory (Yerevan Physics Institute) Foundation (ANSL), State Committee of Science and World Federation of Scientists (WFS), Armenia;
Austrian Academy of Sciences, Austrian Science Fund (FWF): [M 2467-N36] and Nationalstiftung f\"{u}r Forschung, Technologie und Entwicklung, Austria;
Ministry of Communications and High Technologies, National Nuclear Research Center, Azerbaijan;
Conselho Nacional de Desenvolvimento Cient\'{\i}fico e Tecnol\'{o}gico (CNPq), Financiadora de Estudos e Projetos (Finep), Funda\c{c}\~{a}o de Amparo \`{a} Pesquisa do Estado de S\~{a}o Paulo (FAPESP) and Universidade Federal do Rio Grande do Sul (UFRGS), Brazil;
Bulgarian Ministry of Education and Science, within the National Roadmap for Research Infrastructures 2020-2027 (object CERN), Bulgaria;
Ministry of Education of China (MOEC) , Ministry of Science \& Technology of China (MSTC) and National Natural Science Foundation of China (NSFC), China;
Ministry of Science and Education and Croatian Science Foundation, Croatia;
Centro de Aplicaciones Tecnol\'{o}gicas y Desarrollo Nuclear (CEADEN), Cubaenerg\'{\i}a, Cuba;
Ministry of Education, Youth and Sports of the Czech Republic, Czech Republic;
The Danish Council for Independent Research | Natural Sciences, the VILLUM FONDEN and Danish National Research Foundation (DNRF), Denmark;
Helsinki Institute of Physics (HIP), Finland;
Commissariat \`{a} l'Energie Atomique (CEA) and Institut National de Physique Nucl\'{e}aire et de Physique des Particules (IN2P3) and Centre National de la Recherche Scientifique (CNRS), France;
Bundesministerium f\"{u}r Bildung und Forschung (BMBF) and GSI Helmholtzzentrum f\"{u}r Schwerionenforschung GmbH, Germany;
General Secretariat for Research and Technology, Ministry of Education, Research and Religions, Greece;
National Research, Development and Innovation Office, Hungary;
Department of Atomic Energy Government of India (DAE), Department of Science and Technology, Government of India (DST), University Grants Commission, Government of India (UGC) and Council of Scientific and Industrial Research (CSIR), India;
National Research and Innovation Agency - BRIN, Indonesia;
Istituto Nazionale di Fisica Nucleare (INFN), Italy;
Japanese Ministry of Education, Culture, Sports, Science and Technology (MEXT) and Japan Society for the Promotion of Science (JSPS) KAKENHI, Japan;
Consejo Nacional de Ciencia (CONACYT) y Tecnolog\'{i}a, through Fondo de Cooperaci\'{o}n Internacional en Ciencia y Tecnolog\'{i}a (FONCICYT) and Direcci\'{o}n General de Asuntos del Personal Academico (DGAPA), Mexico;
Nederlandse Organisatie voor Wetenschappelijk Onderzoek (NWO), Netherlands;
The Research Council of Norway, Norway;
Pontificia Universidad Cat\'{o}lica del Per\'{u}, Peru;
Ministry of Science and Higher Education, National Science Centre and WUT ID-UB, Poland;
Korea Institute of Science and Technology Information and National Research Foundation of Korea (NRF), Republic of Korea;
Ministry of Education and Scientific Research, Institute of Atomic Physics, Ministry of Research and Innovation and Institute of Atomic Physics and Universitatea Nationala de Stiinta si Tehnologie Politehnica Bucuresti, Romania;
Ministry of Education, Science, Research and Sport of the Slovak Republic, Slovakia;
National Research Foundation of South Africa, South Africa;
Swedish Research Council (VR) and Knut \& Alice Wallenberg Foundation (KAW), Sweden;
European Organization for Nuclear Research, Switzerland;
Suranaree University of Technology (SUT), National Science and Technology Development Agency (NSTDA) and National Science, Research and Innovation Fund (NSRF via PMU-B B05F650021), Thailand;
Turkish Energy, Nuclear and Mineral Research Agency (TENMAK), Turkey;
National Academy of  Sciences of Ukraine, Ukraine;
Science and Technology Facilities Council (STFC), United Kingdom;
National Science Foundation of the United States of America (NSF) and United States Department of Energy, Office of Nuclear Physics (DOE NP), United States of America.
In addition, individual groups or members have received support from:
Czech Science Foundation (grant no. 23-07499S), Czech Republic;
FORTE project, reg.\ no.\ CZ.02.01.01/00/22\_008/0004632, Czech Republic, co-funded by the European Union, Czech Republic;
European Research Council (grant no. 950692), European Union;
ICSC - Centro Nazionale di Ricerca in High Performance Computing, Big Data and Quantum Computing, European Union - NextGenerationEU;
Academy of Finland (Center of Excellence in Quark Matter) (grant nos. 346327, 346328), Finland.

\end{acknowledgement}
\bibliographystyle{utphys}   
\bibliography{bibliography}

\newpage
\appendix
\section{Appendix}
\label{app:update}

This appendix extends the original submission, updating the \ccbar cross sections taking into account recent measurements of the charm fragmentation fractions (FF). The procedure is explained in the following.
With the measured FF in pp collisions at $\sqrt{s} = 5.02$~TeV at the LHC~\cite{ALICE:2021dhb}, the effective branching ratio of charm quarks into electrons can be reevaluated. This first measurement at LHC energies also extends the previous measurements by including the FF of the $\mathrm{\Xi_{c}}^{0}$. For the fragmentation fraction of charm quarks into $\Xi^{+}_{\mathrm{c}}$ baryons, $f(\mathrm{c} \rightarrow \Xi^{+}_{\mathrm{c}})$, the same value as for \mbox{$f(\mathrm{c} \rightarrow \Xi^{0}_{\mathrm c})$} is used, assuming isospin symmetry. This is confirmed by measurements in pp collisions at $\s = 13$~TeV~\cite{ALICE:2023sgl}. The values used are summarised in Table~\ref{tab:summary}. Listed are the fragmentation fractions of c quarks into charm hadrons ($\mathrm{H_{c}}$)~\cite{ALICE:2021dhb}, together with their branching ratios into electrons based on Ref.~\cite{ParticleDataGroup:2024cfk}. In the last column, the product of both is calculated, giving the probability of $\mathrm{c \rightarrow H_{c} \rightarrow e}$ for each charm hadron species. The absolute probability for a c quark to produce an electron after fragmentation and decay is then given by the sum over these values and amounts to $7.33^{+0.61}_{-0.66}$\%. This does not include the contribution of $\rm \Omega_{c}^{0}$, leading to an additional uncertainty of $-0.52\%$. This is added in quadrature to the lower uncertainty resulting in \mbox{$\mathrm{BR(c \to e)} = 7.33^{+0.61}_{-0.84}$},
which corresponds to $0.54^{+0.09}_{-0.12}$\% for the $\ccbar \rightarrow \ee$ case.

\begin{table}[ht!]
    \centering
    \caption{The charm quark fragmentation fractions into charm hadrons $f$(\cHc) from Ref.~\cite{ALICE:2021dhb} are listed together with the respective branching ratios of the hadrons into electron BR(\Hce) from Ref.~\cite{ParticleDataGroup:2024cfk}, as well as their product. The systematic and statistical uncertainties of $f$(\cHc) are added in quadrature, assuming they are uncorrelated.
    If available, the inclusive values for of the branching ratios with electrons in the final state are used. In the case of $\Xi_{\rm{c}}^{+}$ and $\Xi_{\rm{c}}^{0}$ only one decay with an electron in the final state is measured. In these cases, the BR for the respective channel is used.}
    \begin{tabular}{c c c c}
    \toprule
    $\mathrm{H_{c}}$          &  $f\mathrm{(\cHc)}$[\%] & $\mathrm{BR(\Hce)}$[\%] & $f\times\mathrm{BR}$[\%] \\
    \midrule
    $D^{0}$                   & $\mathrm{39.1 \pm 1.7 {\rm{(stat)}} ^{+2.5} _{-2.3} {\rm{(syst)}}}$             & $6.49\pm0.11$       & $2.54^{+ 0.20 }_{- 0.27 }$\\
    $D^{+}$                   & $\mathrm{17.3 \pm 1.8 {\rm{(stat)}} ^{+1.7} _{-1.7} {\rm{(syst)}}}$             & $16.07\pm0.3$       & $2.78^{+ 0.40 }_{- 0.45 }$\\
    $D^{+}_{\mathrm s}$       & $\mathrm{7.3  \pm 1.0 {\rm{(stat)}} ^{+1.9} _{-0.1} {\rm{(syst)}}}$             & $6.33\pm0.15$       & $0.46^{+ 0.14 }_{- 0.09 }$\\
    $\Lambda^{+}_{\mathrm c}$ & $\mathrm{20.4 \pm 1.3 {\rm{(stat)}} ^{+1.6} _{-1.6} {\rm{(syst)}}}$             & $4.06\pm0.13$       & $0.83^{+ 0.09 }_{- 0.09 }$\\
    $\Xi^{0}_{\mathrm c}$     & $\mathrm{8.0  \pm 1.2 {\rm{(stat)}} ^{+2.5} _{-2.3} {\rm{(syst)}}}$             & $1.94\pm0.55$  & $0.16^{+ 0.05 }_{- 0.05}$\\
    $\Xi^{+}_{\mathrm c}$     & $\mathrm{8.0  \pm 1.2 {\rm{(stat)}} ^{+2.5} _{-2.3} {\rm{(syst)}}}$             & $7\pm4$             & $0.56^{+ 0.37 }_{- 0.37 }$\\
    \cmidrule{4-4}
                              &                                               &                     & $7.33^{+0.61}_{-0.66}$ \\
    \bottomrule 
    &
    \end{tabular}
    \label{tab:summary}
\end{table}
The comparison with the previously used value of $\mathrm{BR(c \to e)}=9.6\%$ leads to a larger \ccbar production cross section at midrapidity ($\mathrm{d}\sigma_{\ccbar}/\mathrm{d}y|_{y=0}$) by a factor 1.72 as the change in $\mathrm{BR(c \to e)}$ enters the final results squared. The previously assigned uncertainty on BR($\ccbar\rightarrow\ee$) was 22\%, which is now reevaluated  to $^{+23}_{-17}$\%. 
The \ccbar cross sections in pp collisions at $\sqrt{s} = 7$ and 13 TeV~\cite{ALICE:2018fvj, ALICE:2018gev} can be updated with the same procedure.
The reevaluated \ccbar production cross sections at midrapidity are summarised in Table~\ref{tab:ccCrosssectionsNew}. Note that with respect to Refs.~\cite{ALICE:2018fvj, ALICE:2018gev, ALICE:2020mfy} only the \ccbar cross sections change.

\begin{table}[h]
    \centering
    \caption{\ccbar production cross sections at midrapidity extracted at different collision energies using the event generators PYTHIA~6 and POWHEG with an effective branching ratio of charm quarks to electrons based on fragmentation fractions measured by the ALICE collaboration~\cite{ALICE:2021dhb}.}
    \begin{tabular}{c c c}
    \toprule
    $\sqrt{s}$ (TeV) & \multicolumn{2}{c}{ $\sigma_{\mathrm {c\bar{c}}}|_{y=0}$ ($\mu$b)}  \\
    \midrule
    & {PYTHIA} & {POWHEG}  \\
    5.02 & $\rm 900 \pm 105(stat) \pm 45(syst)$ $^{+208}_{-152}$ (BR) & $\rm 1299 \pm 137(stat) \pm 65(syst)$ $^{+300}_{-220}$ (BR) \\
    7 & $\rm 1468 \pm 211(stat) \pm 249(syst)$ $^{+339}_{-248}$ (BR)  & $\rm 2150 \pm 266(stat) \pm 366(syst)$ $^{+497}_{-363}$ (BR)  \\
    13 & $\rm 1674 \pm 237(stat) \pm 241(syst)$ $^{+387}_{-283}$ (BR) & $\rm 2435 \pm 316(stat) \pm 351(syst)$ $^{+562}_{-411}$ (BR)  \\
 \bottomrule
  &
    \end{tabular}
    \label{tab:ccCrosssectionsNew}
\end{table}


It was seen that using a more recent implementation of the PYTHIA event generator that includes beyond-leading-colour approximation (mode 2) the extracted cross section is not changed~\cite{Bierlich:2022pfr, Christiansen:2015yqa}.


As for the beauty sector, enhanced baryon-to-meson ratios for beauty hadrons~\cite{LHCb:2019fns} and non-prompt charm hadrons~\cite{ALICE:2023wbx} have been measured in hadronic collisions at the LHC compared to results in e$^{+}$e$^{-}$ collisions at lower energies. Since the semi-leptonic branching ratios of the different beauty hadrons are very close to each other, these observations affect only the dielectron pairs involving at least one charm hadron decay ($\mathrm{b \to c \to e}$). Since no decays were forced in the creation of the templates, the fragmentation and decay in the relevant PYTHIA version are responsible for reproducing the measurements. A reweighting was performed to study the validity of the utilized templates to extract the cross-sections. It was found that the overall effect on the shape of the beauty template is negligible in the phase space where the measurement is sensitive to the beauty production.
%
%

\section{The ALICE Collaboration}
\label{app:collab}
\begin{flushleft} 
\small

S.~Acharya\,\orcidlink{0000-0002-9213-5329}\,$^{\rm 127}$, 
A.~Adler, 
J.~Adolfsson\,\orcidlink{0000-0001-5651-4025}\,, 
A.~Agarwal$^{\rm 135}$, 
G.~Aglieri Rinella\,\orcidlink{0000-0002-9611-3696}\,$^{\rm 32}$, 
L.~Aglietta\,\orcidlink{0009-0003-0763-6802}\,$^{\rm 24}$, 
M.~Agnello\,\orcidlink{0000-0002-0760-5075}\,$^{\rm 29}$, 
N.~Agrawal\,\orcidlink{0000-0003-0348-9836}\,$^{\rm 25}$, 
Z.~Ahammed\,\orcidlink{0000-0001-5241-7412}\,$^{\rm 135}$, 
S.~Ahmad\,\orcidlink{0000-0003-0497-5705}\,$^{\rm 15}$
% ...

\section*{Affiliation Notes}

$^{\rm I}$ Deceased\\
$^{\rm II}$ Also at: Max-Planck-Institut fur Physik, Munich, Germany\\
$^{\rm III}$ Also at: Czech Technical University in Prague (CZ)\\
$^{\rm IV}$ Also at: Italian National Agency for New Technologies, Energy and Sustainable Economic Development (ENEA), Bologna, Italy\\
$^{\rm V}$ Also at: Dipartimento DET del Politecnico di Torino, Turin, Italy\\
$^{\rm VI}$ Also at: Lehigh University, United States\\
$^{\rm VII}$ Also at: Department of Applied Physics, Aligarh Muslim University, Aligarh, India\\
$^{\rm VIII}$ Also at: Institute of Theoretical Physics, University of Wroclaw, Poland\\
$^{\rm IX}$ Also at: Facultad de Ciencias, Universidad Nacional Aut\'{o}noma de M\'{e}xico, Mexico City, Mexico\\

\section*{Collaboration Institutes}

$^{1}$ A.I. Alikhanyan National Science Laboratory (Yerevan Physics Institute) Foundation, Yerevan, Armenia\\
$^{2}$ AGH University of Krakow, Cracow, Poland\\
$^{3}$ Bogolyubov Institute for Theoretical Physics, National Academy of Sciences of Ukraine, Kyiv, Ukraine\\
$^{4}$ Bose Institute, Department of Physics  and Centre for Astroparticle Physics and Space Science (CAPSS), Kolkata, India\\
$^{5}$ California Polytechnic State University, San Luis Obispo, California, United States\\
$^{6}$ Central China Normal University, Wuhan, China\\
$^{7}$ Centro de Aplicaciones Tecnol\'{o}gicas y Desarrollo Nuclear (CEADEN), Havana, Cuba\\
$^{8}$ Centro de Investigaci\'{o}n y de Estudios Avanzados (CINVESTAV), Mexico City and M\'{e}rida, Mexico\\
$^{9}$ Chicago State University, Chicago, Illinois, United States\\
$^{10}$ China Nuclear Data Center, China Institute of Atomic Energy, Beijing, China\\
$^{11}$ China University of Geosciences, Wuhan, China\\
$^{12}$ Chungbuk National University, Cheongju, Republic of Korea\\
$^{13}$ Comenius University Bratislava, Faculty of Mathematics, Physics and Informatics, Bratislava, Slovak Republic\\
$^{14}$ Creighton University, Omaha, Nebraska, United States\\
$^{15}$ Department of Physics, Aligarh Muslim University, Aligarh, India\\
$^{16}$ Department of Physics, Pusan National University, Pusan, Republic of Korea\\
$^{17}$ Department of Physics, Sejong University, Seoul, Republic of Korea\\
$^{18}$ Department of Physics, University of California, Berkeley, California, United States\\
$^{19}$ Department of Physics, University of Oslo, Oslo, Norway\\
$^{20}$ Department of Physics and Technology, University of Bergen, Bergen, Norway\\
$^{21}$ Dipartimento di Fisica, Universit\`{a} di Pavia, Pavia, Italy\\
$^{22}$ Dipartimento di Fisica dell'Universit\`{a} and Sezione INFN, Cagliari, Italy\\
$^{23}$ Dipartimento di Fisica dell'Universit\`{a} and Sezione INFN, Trieste, Italy\\
$^{24}$ Dipartimento di Fisica dell'Universit\`{a} and Sezione INFN, Turin, Italy\\
$^{25}$ Dipartimento di Fisica e Astronomia dell'Universit\`{a} and Sezione INFN, Bologna, Italy\\
$^{26}$ Dipartimento di Fisica e Astronomia dell'Universit\`{a} and Sezione INFN, Catania, Italy\\
$^{27}$ Dipartimento di Fisica e Astronomia dell'Universit\`{a} and Sezione INFN, Padova, Italy\\
$^{28}$ Dipartimento di Fisica `E.R.~Caianiello' dell'Universit\`{a} and Gruppo Collegato INFN, Salerno, Italy\\
$^{29}$ Dipartimento DISAT del Politecnico and Sezione INFN, Turin, Italy\\
$^{30}$ Dipartimento di Scienze MIFT, Universit\`{a} di Messina, Messina, Italy\\
$^{31}$ Dipartimento Interateneo di Fisica `M.~Merlin' and Sezione INFN, Bari, Italy\\
$^{32}$ European Organization for Nuclear Research (CERN), Geneva, Switzerland\\
$^{33}$ Faculty of Electrical Engineering, Mechanical Engineering and Naval Architecture, University of Split, Split, Croatia\\
$^{34}$ Faculty of Nuclear Sciences and Physical Engineering, Czech Technical University in Prague, Prague, Czech Republic\\
$^{35}$ Faculty of Physics, Sofia University, Sofia, Bulgaria\\
$^{36}$ Faculty of Science, P.J.~\v{S}af\'{a}rik University, Ko\v{s}ice, Slovak Republic\\
$^{37}$ Faculty of Technology, Environmental and Social Sciences, Bergen, Norway\\
$^{38}$ Frankfurt Institute for Advanced Studies, Johann Wolfgang Goethe-Universit\"{a}t Frankfurt, Frankfurt, Germany\\
$^{39}$ Fudan University, Shanghai, China\\
$^{40}$ Gangneung-Wonju National University, Gangneung, Republic of Korea\\
$^{41}$ Gauhati University, Department of Physics, Guwahati, India\\
$^{42}$ Helmholtz-Institut f\"{u}r Strahlen- und Kernphysik, Rheinische Friedrich-Wilhelms-Universit\"{a}t Bonn, Bonn, Germany\\
$^{43}$ Helsinki Institute of Physics (HIP), Helsinki, Finland\\
$^{44}$ High Energy Physics Group,  Universidad Aut\'{o}noma de Puebla, Puebla, Mexico\\
$^{45}$ Horia Hulubei National Institute of Physics and Nuclear Engineering, Bucharest, Romania\\
$^{46}$ HUN-REN Wigner Research Centre for Physics, Budapest, Hungary\\
$^{47}$ Indian Institute of Technology Bombay (IIT), Mumbai, India\\
$^{48}$ Indian Institute of Technology Indore, Indore, India\\
$^{49}$ INFN, Laboratori Nazionali di Frascati, Frascati, Italy\\
$^{50}$ INFN, Sezione di Bari, Bari, Italy\\
$^{51}$ INFN, Sezione di Bologna, Bologna, Italy\\
$^{52}$ INFN, Sezione di Cagliari, Cagliari, Italy\\
$^{53}$ INFN, Sezione di Catania, Catania, Italy\\
$^{54}$ INFN, Sezione di Padova, Padova, Italy\\
$^{55}$ INFN, Sezione di Pavia, Pavia, Italy\\
$^{56}$ INFN, Sezione di Torino, Turin, Italy\\
$^{57}$ INFN, Sezione di Trieste, Trieste, Italy\\
$^{58}$ Inha University, Incheon, Republic of Korea\\
$^{59}$ Institute for Gravitational and Subatomic Physics (GRASP), Utrecht University/Nikhef, Utrecht, Netherlands\\
$^{60}$ Institute of Experimental Physics, Slovak Academy of Sciences, Ko\v{s}ice, Slovak Republic\\
$^{61}$ Institute of Physics, Homi Bhabha National Institute, Bhubaneswar, India\\
$^{62}$ Institute of Physics of the Czech Academy of Sciences, Prague, Czech Republic\\
$^{63}$ Institute of Space Science (ISS), Bucharest, Romania\\
$^{64}$ Institut f\"{u}r Kernphysik, Johann Wolfgang Goethe-Universit\"{a}t Frankfurt, Frankfurt, Germany\\
$^{65}$ Instituto de Ciencias Nucleares, Universidad Nacional Aut\'{o}noma de M\'{e}xico, Mexico City, Mexico\\
$^{66}$ Instituto de F\'{i}sica, Universidade Federal do Rio Grande do Sul (UFRGS), Porto Alegre, Brazil\\
$^{67}$ Instituto de F\'{\i}sica, Universidad Nacional Aut\'{o}noma de M\'{e}xico, Mexico City, Mexico\\
$^{68}$ iThemba LABS, National Research Foundation, Somerset West, South Africa\\
$^{69}$ Jeonbuk National University, Jeonju, Republic of Korea\\
$^{70}$ Johann-Wolfgang-Goethe Universit\"{a}t Frankfurt Institut f\"{u}r Informatik, Fachbereich Informatik und Mathematik, Frankfurt, Germany\\
$^{71}$ Korea Institute of Science and Technology Information, Daejeon, Republic of Korea\\
$^{72}$ KTO Karatay University, Konya, Turkey\\
$^{73}$ Laboratoire de Physique Subatomique et de Cosmologie, Universit\'{e} Grenoble-Alpes, CNRS-IN2P3, Grenoble, France\\
$^{74}$ Lawrence Berkeley National Laboratory, Berkeley, California, United States\\
$^{75}$ Lund University Department of Physics, Division of Particle Physics, Lund, Sweden\\
$^{76}$ Marietta Blau Institute, Vienna, Austria\\
$^{77}$ Nagasaki Institute of Applied Science, Nagasaki, Japan\\
$^{78}$ Nara Women{'}s University (NWU), Nara, Japan\\
$^{79}$ National and Kapodistrian University of Athens, School of Science, Department of Physics , Athens, Greece\\
$^{80}$ National Centre for Nuclear Research, Warsaw, Poland\\
$^{81}$ National Institute of Science Education and Research, Homi Bhabha National Institute, Jatni, India\\
$^{82}$ National Nuclear Research Center, Baku, Azerbaijan\\
$^{83}$ National Research and Innovation Agency - BRIN, Jakarta, Indonesia\\
$^{84}$ Niels Bohr Institute, University of Copenhagen, Copenhagen, Denmark\\
$^{85}$ Nikhef, National institute for subatomic physics, Amsterdam, Netherlands\\
$^{86}$ Nuclear Physics Group, STFC Daresbury Laboratory, Daresbury, United Kingdom\\
$^{87}$ Nuclear Physics Institute of the Czech Academy of Sciences, Husinec-\v{R}e\v{z}, Czech Republic\\
$^{88}$ Oak Ridge National Laboratory, Oak Ridge, Tennessee, United States\\
$^{89}$ Ohio State University, Columbus, Ohio, United States\\
$^{90}$ Physics department, Faculty of science, University of Zagreb, Zagreb, Croatia\\
$^{91}$ Physics Department, Panjab University, Chandigarh, India\\
$^{92}$ Physics Department, University of Jammu, Jammu, India\\
$^{93}$ Physics Program and International Institute for Sustainability with Knotted Chiral Meta Matter (WPI-SKCM$^{2}$), Hiroshima University, Hiroshima, Japan\\
$^{94}$ Physikalisches Institut, Eberhard-Karls-Universit\"{a}t T\"{u}bingen, T\"{u}bingen, Germany\\
$^{95}$ Physikalisches Institut, Ruprecht-Karls-Universit\"{a}t Heidelberg, Heidelberg, Germany\\
$^{96}$ Physik Department, Technische Universit\"{a}t M\"{u}nchen, Munich, Germany\\
$^{97}$ Politecnico di Bari and Sezione INFN, Bari, Italy\\
$^{98}$ Research Division and ExtreMe Matter Institute EMMI, GSI Helmholtzzentrum f\"ur Schwerionenforschung GmbH, Darmstadt, Germany\\
$^{99}$ Saga University, Saga, Japan\\
$^{100}$ Saha Institute of Nuclear Physics, Homi Bhabha National Institute, Kolkata, India\\
$^{101}$ School of Physics and Astronomy, University of Birmingham, Birmingham, United Kingdom\\
$^{102}$ Secci\'{o}n F\'{\i}sica, Departamento de Ciencias, Pontificia Universidad Cat\'{o}lica del Per\'{u}, Lima, Peru\\
$^{103}$ SUBATECH, IMT Atlantique, Nantes Universit\'{e}, CNRS-IN2P3, Nantes, France\\
$^{104}$ Sungkyunkwan University, Suwon City, Republic of Korea\\
$^{105}$ Suranaree University of Technology, Nakhon Ratchasima, Thailand\\
$^{106}$ Technical University of Ko\v{s}ice, Ko\v{s}ice, Slovak Republic\\
$^{107}$ The Henryk Niewodniczanski Institute of Nuclear Physics, Polish Academy of Sciences, Cracow, Poland\\
$^{108}$ The University of Texas at Austin, Austin, Texas, United States\\
$^{109}$ Universidad Aut\'{o}noma de Sinaloa, Culiac\'{a}n, Mexico\\
$^{110}$ Universidade de S\~{a}o Paulo (USP), S\~{a}o Paulo, Brazil\\
$^{111}$ Universidade Estadual de Campinas (UNICAMP), Campinas, Brazil\\
$^{112}$ Universidade Federal do ABC, Santo Andre, Brazil\\
$^{113}$ Universitatea Nationala de Stiinta si Tehnologie Politehnica Bucuresti, Bucharest, Romania\\
$^{114}$ University of Cape Town, Cape Town, South Africa\\
$^{115}$ University of Derby, Derby, United Kingdom\\
$^{116}$ University of Houston, Houston, Texas, United States\\
$^{117}$ University of Jyv\"{a}skyl\"{a}, Jyv\"{a}skyl\"{a}, Finland\\
$^{118}$ University of Kansas, Lawrence, Kansas, United States\\
$^{119}$ University of Liverpool, Liverpool, United Kingdom\\
$^{120}$ University of Science and Technology of China, Hefei, China\\
$^{121}$ University of South-Eastern Norway, Kongsberg, Norway\\
$^{122}$ University of Tennessee, Knoxville, Tennessee, United States\\
$^{123}$ University of the Witwatersrand, Johannesburg, South Africa\\
$^{124}$ University of Tokyo, Tokyo, Japan\\
$^{125}$ University of Tsukuba, Tsukuba, Japan\\
$^{126}$ Universit\"{a}t M\"{u}nster, Institut f\"{u}r Kernphysik, M\"{u}nster, Germany\\
$^{127}$ Universit\'{e} Clermont Auvergne, CNRS/IN2P3, LPC, Clermont-Ferrand, France\\
$^{128}$ Universit\'{e} de Lyon, CNRS/IN2P3, Institut de Physique des 2 Infinis de Lyon, Lyon, France\\
$^{129}$ Universit\'{e} de Strasbourg, CNRS, IPHC UMR 7178, F-67000 Strasbourg, France, Strasbourg, France\\
$^{130}$ Universit\'{e} Paris-Saclay, Centre d'Etudes de Saclay (CEA), IRFU, D\'{e}partment de Physique Nucl\'{e}aire (DPhN), Saclay, France\\
$^{131}$ Universit\'{e}  Paris-Saclay, CNRS/IN2P3, IJCLab, Orsay, France\\
$^{132}$ Universit\`{a} degli Studi di Foggia, Foggia, Italy\\
$^{133}$ Universit\`{a} del Piemonte Orientale, Vercelli, Italy\\
$^{134}$ Universit\`{a} di Brescia, Brescia, Italy\\
$^{135}$ Variable Energy Cyclotron Centre, Homi Bhabha National Institute, Kolkata, India\\
$^{136}$ Warsaw University of Technology, Warsaw, Poland\\
$^{137}$ Wayne State University, Detroit, Michigan, United States\\
$^{138}$ Yale University, New Haven, Connecticut, United States\\
$^{139}$ Yonsei University, Seoul, Republic of Korea\\
$^{140}$  Zentrum  f\"{u}r Technologie und Transfer (ZTT), Worms, Germany\\
$^{141}$ Affiliated with an institute formerly covered by a cooperation agreement with CERN\\
$^{142}$ Affiliated with an international laboratory covered by a cooperation agreement with CERN.\\

\end{flushleft} 

\end{document}